\documentclass[%
 reprint,
superscriptaddress,
 amsmath,amssymb,
 aps,
 pra,
floatfix,
twocolumn,
]{revtex4-2}

\usepackage{diagbox}
\usepackage{array}
\usepackage{graphicx}% Include figure files
\usepackage{dcolumn}% Align table columns on decimal point
\usepackage{comment}
\usepackage{subfigure}
\usepackage[caption=false]{subfig}
\usepackage{bm}% bold math
\begin{document}

\preprint{APS/123-QED}

\title{ On-demand generation of all four Bell states using a single PPKTP entangled photon source}

\author{Gayatri Thik}
\affiliation{School of Quantum Technology, Defence Institute of Advanced Technology, Pune, India.}

\author{Amit Loyal}
\altaffiliation{Present affiliation: College of Optics and Photonics (CREOL), University of Central Florida, Orlando, Florida, USA.}
\affiliation{School of Quantum Technology, Defence Institute of Advanced Technology, Pune, India.}

\author{Srinivasan K}
\email{srinik@diat.ac.in}
\affiliation{School of Quantum Technology, Defence Institute of Advanced Technology, Pune, India.}

\author{Raghavan G}
\email{graghavan@diat.ac.in}
\affiliation{School of Quantum Technology, Defence Institute of Advanced Technology, Pune, India.}

\begin{abstract}
We present a compact, automated, high-brightness entangled photon source capable of generating all four Bell states with high fidelity. The system utilizes a type-0 quasi-phase-matched PPKTP crystal embedded within a polarization Sagnac interferometer. We introduce a switching scheme based on the controlled, motorized translation of the nonlinear crystal. This device is capable of generating any one of the Bell states on-demand. Experimentally, we demonstrate that translating the crystal from the interferometer's balanced position repeatedly toggles the state between $|\phi^+ \rangle$ and $|\phi^- \rangle$ (as well as $|\psi^+ \rangle$ and $|\psi^- \rangle$) at regular intervals of $122 \pm 14 ~\mu m$. Subsequently, a half-wave plate (HWP) in the idler arm transitions between the quantum states $|\phi^{\pm}\rangle$ and $|\psi^{\pm}\rangle$. While the non-collinear geometry imposes an upper limit on the translation range as verified via EMCCD imaging, the source however, displays very little change of intensity in the operational window. State purity and entanglement are certified through quantum state tomography (QST), visibility measurements, Bell state measurements (BSM), and CHSH inequality violations, confirming that the source is robust and provides a repeatable, high-fidelity output.
\end{abstract}

\maketitle

%\tableofcontents

\section{\label{sec:level1}Introduction}

Maximally entangled bipartite quantum states, commonly known as Bell states, are fundamental resources in quantum technologies \cite{Bennett_PhysRevLett.70.1895,Ekert1991,Nielsen_Chuang_2010}. These states play a crucial role in several quantum information protocols such as two-photon interference experiments, entanglement swapping, measurement-device-independent quantum key distribution (MDI-QKD), and quantum teleportation \cite{HOM_PhysRevLett.59.2044, Pan_Entang_swap_PhysRevLett.80.3891,MDI-QKD_PhysRevLett.108.130503,Bennett_PhysRevLett.70.1895}. An important property of Bell states is that any arbitrary bipartite quantum state can be expressed in terms of the Bell basis \cite{Nielsen_Chuang_2010}.

Optical Bell states can be generated using four-wave mixing or spontaneous parametric down-conversion (SPDC) \cite{Shih1988,anwarAli_10.1063/5.0023103,Guo_2017}. Among these methods, SPDC remains the most widely used technique for generating entangled photon pairs \cite{SPDC_PhysRevLett.75.4337}. Considerable effort has been devoted to developing high-brightness and high-quality entangled photon sources using nonlinear crystals \cite{Kwait_99_PhysRevA.60.R773,Marco_F_Fiorentino:07,Konig_10.1063/1.1668320}.

Several nonlinear crystals such as BBO, BiBO, PPLN, and PPKTP have been used to generate entangled photons \cite{BBO_SansaPerna2022,BiBO_Rangarajan2009,Jabir2017,PPLN_Szlachetka2023}. Among these materials, periodically poled potassium titanyl phosphate (PPKTP) and periodically poled lithium niobate (PPLN) are widely used for developing high-brightness entangled photon sources. Although PPLN crystals can provide high spectral brightness, they are extremely sensitive to temperature fluctuations and typically require heating above ambient temperature for stable operation \cite{Tanzilli}.

In contrast, type-0 quasi-phase-matched PPKTP crystals can generate entangled photons with high spectral brightness at room temperature, making them highly suitable for practical quantum optical systems \cite{Konig_10.1063/1.1668320, Kim_PhysRevA.73.012316}. These crystals can be used in both single-pass linear configurations and polarization Sagnac interferometer geometries \cite{PPKTP_linear_Guo2023,PPKTP_linear_Park2025,sagnac_Cai2022,Jabir2017}.
Between these two approaches, polarization Sagnac interferometer-based sources offer high brightness and better phase stability \cite{Kim_PhysRevA.73.012316, Jin2014}. Such sources can be configured to produce any one of the four Bell states given by \cite{Nielsen_Chuang_2010, Gisin1998, Kim2019}:

\begin{equation}
\begin{aligned}
|\phi^{+}\rangle &= \frac{1}{\sqrt{2}}\left(|HH\rangle + |VV\rangle\right) \\
|\phi^{-}\rangle &= \frac{1}{\sqrt{2}}\left(|HH\rangle - |VV\rangle\right) \\
|\psi^{+}\rangle &= \frac{1}{\sqrt{2}}\left(|HV\rangle + |VH\rangle\right) \\
|\psi^{-}\rangle &= \frac{1}{\sqrt{2}}\left(|HV\rangle - |VH\rangle\right)
\label{state}
\end{aligned}
\end{equation}
However, in most conventional configurations, the system is designed to produce only one specific Bell state. Generation of the other Bell states requires additional optical components such as quarter-wave plates (QWPs) and half-wave plates (HWPs) \cite{Sundar:25_VariableBellState, Mishra2024, Kim2019}. Therefore, the ability to generate all four Bell states using minimal modifications in a single experimental setup is highly desirable. In earlier work, it has been observed that compensation of phase offset is achieved by the translation of the nonlinear crystal, while the generation of the remaining Bell states is accomplished via local unitary transformations implemented using HWP and QWP \cite{Kim2019}. The present work focuses on developing an automated, single tunable polarization-entangled photon source capable of switching between all four Bell states with minimal additional components. A single tunable source with these capabilities is important for developing a cost-effective source with reduced optical complexity and minimal alignment requirements. These sources easily switch between Bell states with similar experimental conditions for all four Bell states. We adopt the crystal translation approach to achieve this capability based on the insight and observations that the PPKTP crystal placed at the balanced position of the Sagnac interferometer 
produces zero phase difference between the two counter-propagating orthogonally 
polarized photon pairs $|H_s\rangle \otimes |H_i\rangle$ and $|V_s\rangle \otimes |V_i\rangle$, resulting in maximally entangled states. When the crystal is translated from its central position, a relative phase ($\phi$) in direct proportion  to the crystal displacement ($x$) is introduced between the counter propagating beams. When the displacement is small this phase changes approximately as:

\begin{equation}
    \phi(x) = \frac{2\pi}{r} x + \phi_0
\end{equation}
Here,  $x$ represents the crystal displacement from the balanced central position ($x=0$) and $r$ is the spatial interference period, which is determined by the pump wavelength and the refractive indices of the crystal at the respective wavelengths. $\phi_0$ is the initial phase at the balanced position, where $\phi_0 = 0$ corresponds to the generation of the $|\phi^{+}\rangle$ state.
%\end{itemize}

As the crystal is continuously translated, the relative phase $\phi$ evolves, causing the quantum state to switch periodically between $|\phi^{+}\rangle$ and $|\phi^{-}\rangle$ at regular spatial intervals. Continuously translating the crystal from the balanced position switches between $|\phi^{+}\rangle$ to $|\phi^{-}\rangle$ and $|\phi^{-}\rangle$ to $|\phi^{+}\rangle$ at a regular interval. Furthermore, by inserting a HWP oriented at $45^\circ$ in the idler arm, the remaining Bell states $|\psi^+\rangle$ and $|\psi^-\rangle $ can also be generated through crystal translation. Thus, the proposed scheme enables the generation of all four Bell states using a single entangled photon source with only crystal translation and one HWP.

The Bell states thus generated are certified in three stages. First, the states are projected onto the diagonal polarization basis by placing polarizers in the signal and idler arms while translating the crystal from its balanced position. This step facilitates the easy certification of the Bell states without taking recourse to quantum state tomography requiring sixteen projective measurements.  Second, Bell state measurements (BSMs) are performed to confirm the generated states and are consistent with expected results. The Bell state analyzer test is one of the major building blocks for quantum technology, as it plays a major part in entanglement swapping, quantum teleportation, quantum repeaters and photonic quantum computing \cite{bellStateMeasurement,teleportation_DAurelio,Kim_PhysRevLett.86.1370}. The Bell state analyzer test is performed by distributing the photons with an optical fiber followed by the Bell state measurement setup containing a beam splitter (BS) and a pair of polarizing beam splitters (PBS) in each output mode of the BS. Finally, complete quantum state tomography is carried out to fully characterize the entangled states. 

The work reported here is organized as follows.
Section II presents the theoretical framework for verifying the generated Bell states through projection in the diagonal basis and the corresponding experimental results. Section III explains the Bell state measurement technique used for verification. Section IV details the experimental setup. Section V discusses the experimental results and analysis. Finally, Section VI presents the concluding remarks.

%Section III describes the generation of all four Bell states using crystal translation in the Sagnac interferometer.

\section{Projection of Bell's states on the diagonal basis}
In order to certify the quality of the Bell states generated through crystal translation, the PPKTP crystal is placed at the balanced position of the interferometer to generate the $|\phi^+\rangle$ state. The translation of the crystal  from the balanced position of the Sagnac interferometer introduces a phase difference between the $|HH\rangle$ and $|VV\rangle$ components. The crystal translation is done with a linear translation stage fitted with  a piezoelectric motor. This enables the generation of orthogonal Bell states $|\phi^+\rangle$ and $|\phi^-\rangle$ in a controlled and repeatable manner. This can be confirmed by performing the quantum state tomography. However, this requires $16$ projective measurements. This section is dedicated to the verification of the generated Bell states just by measuring the visibility in the $H-V$ basis, followed by projecting the system on the $|++\rangle$ state by placing two polarizers at angle $+45^\circ$. The Bell state measurement is performed on the generated quantum states to characterize them further. 

Let us assume that an entangled photon source emits photon pairs, and it is guaranteed to be one of the states,
\begin{equation}
    |\phi\rangle = \frac{1}{\sqrt{2}} \left(|H_sH_i\rangle + e^{i \phi} |V_sV_i\rangle\right)
\end{equation}

and 
\begin{equation}
    |\psi\rangle = \frac{1}{\sqrt{2}} \left(|H_sV_i\rangle + e^{i \phi} |V_sH_i\rangle\right)
\end{equation}

\begin{figure}
    \centering
    \includegraphics[width=1\linewidth]{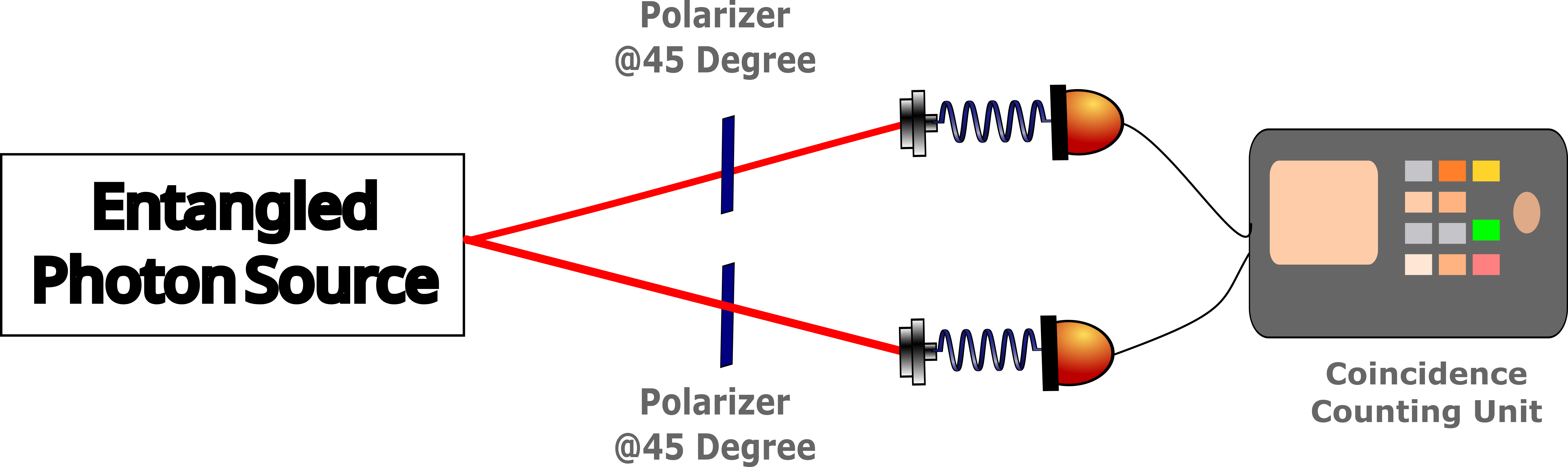} \caption{Projection of the entangled photon pairs in the $|+_s +_i\rangle$} state.
    \label{fig:blackbox}
\end{figure}

Firstly, with the visibility in the H-V basis, one can confirm between the $|\phi\rangle$ and $|\psi\rangle$ states. Further, these photon pairs are projected onto the diagonal basis state by keeping two polarizers at $+ 45^\circ$ in both signal and idler arms as given in the Fig. \ref{fig:blackbox}. On translating the crystal in the Sagnac interferometer, the phase difference $\phi$ continuously changes from $0$ to $2n\pi$, where $n$ is an integer. The probability of measuring the quantum states $|\phi\rangle$ and $|\psi\rangle$ onto the diagonal basis state $|+_s+_i\rangle$ results in,
\begin{equation}
    P(|+_s+_i\rangle) = \frac{1}{4}\left(1 + \cos \phi \right)
\end{equation}

For the state $|\phi\rangle$, continuously changing the relative phase $\phi$ from $0$, results in the generation of $|\phi^+\rangle$ state whenever $\phi = 2n \pi$ and $|\phi^-\rangle$ state whenever $\phi = (2n+1) \pi$. Similarly, for the quantum state $|\psi\rangle$ changing the relative phase $\phi$ from $0$, results in $|\psi^+\rangle$ state whenever $\phi = 2n \pi$ and $|\psi^-\rangle$ state whenever $\phi = (2n+1) \pi$ . 

The coincidence count measurement gives us the probability $P(|+_s +_i\rangle)$ of finding the system in the $|+_s +_i\rangle$ state. The probability of finding Bell states $|\phi^+ \rangle$ and $|\psi^+ \rangle$ in the $|+_s +_i\rangle$ state is given by,
\begin{equation}
    P(|+_s +_i\rangle) = \frac{1}{2}
\end{equation}
Similarly the probability $P(|+_s +_i\rangle)$ of finding the Bell states $|\phi^- \rangle$ and $|\psi^- \rangle$ in the $|+_s +_i\rangle$ state is given by,
\begin{equation}
    P(|+_s +_i\rangle) = 0
\end{equation}

\begin{figure}
    \centering
    \includegraphics[width=1\linewidth]{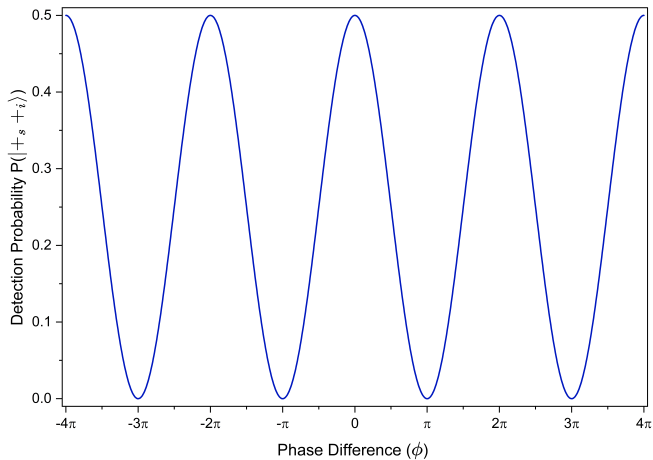}
    \caption{Simulation of detection probability $P(|+_s +_i\rangle)$ for quantum states $|\phi\rangle$ and $|\psi\rangle$ for different values of $\phi$.}
    \label{fig:simulated cosine squre}
\end{figure}

Fig.~\ref{fig:simulated cosine squre} provides the simulation results of detection probability $P(|+_s +_i\rangle)$ for quantum states $|\phi\rangle$ and $|\psi\rangle$ for different values of $\phi$. 

The theoretical model predicts a specific behavior for the coincidence count rate ($C_{DD}$) as the crystal is translated. Here, $C_{DD}$ is the coincidence count rate when the downconverted photons are projected onto the diagonal state $|+_s +_i \rangle$. According to the relationship $C_{DD} \propto \frac{1}{2}(1 + \cos\phi)$, the following states are identified based on the observed patterns in the correlation curve:

\begin{itemize}
    \item \textbf{Maxima} in $C_{DD}$ occur when the relative phase satisfies $\phi = 2n\pi$ (where $n$ is an integer), which identifies the generated state as $|\phi^{+}\rangle$ ( or $|\psi^{+}\rangle$).
    \item \textbf{Minima} in $C_{DD}$ occur when the phase satisfies $\phi = (2n+1)\pi$, identifying the state as $|\phi^{-}\rangle$ (or $|\psi^{-}\rangle$).
\end{itemize}    

\begin{figure}
    \centering
    \includegraphics[width=1\linewidth]{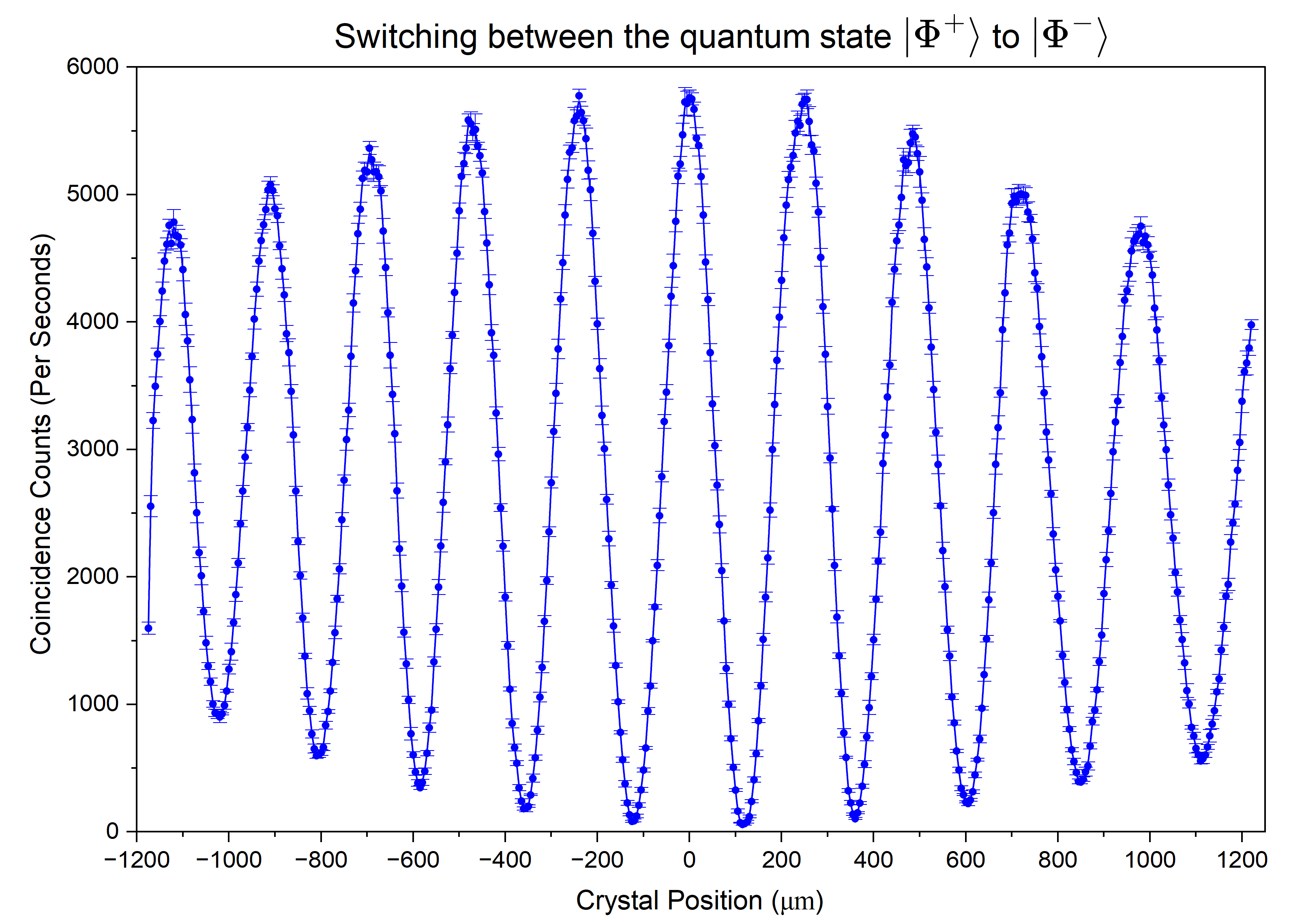}
    \caption{Projection of Bell state $|\phi\rangle$ onto the $|+_s+_i\rangle$ basis state. The balanced position of the crystal is considered to be zero.}
    \label{fig:Switching_between_the_quantum_state_Phi}
\end{figure}

\begin{figure}
    \centering
    \includegraphics[width=1\linewidth]{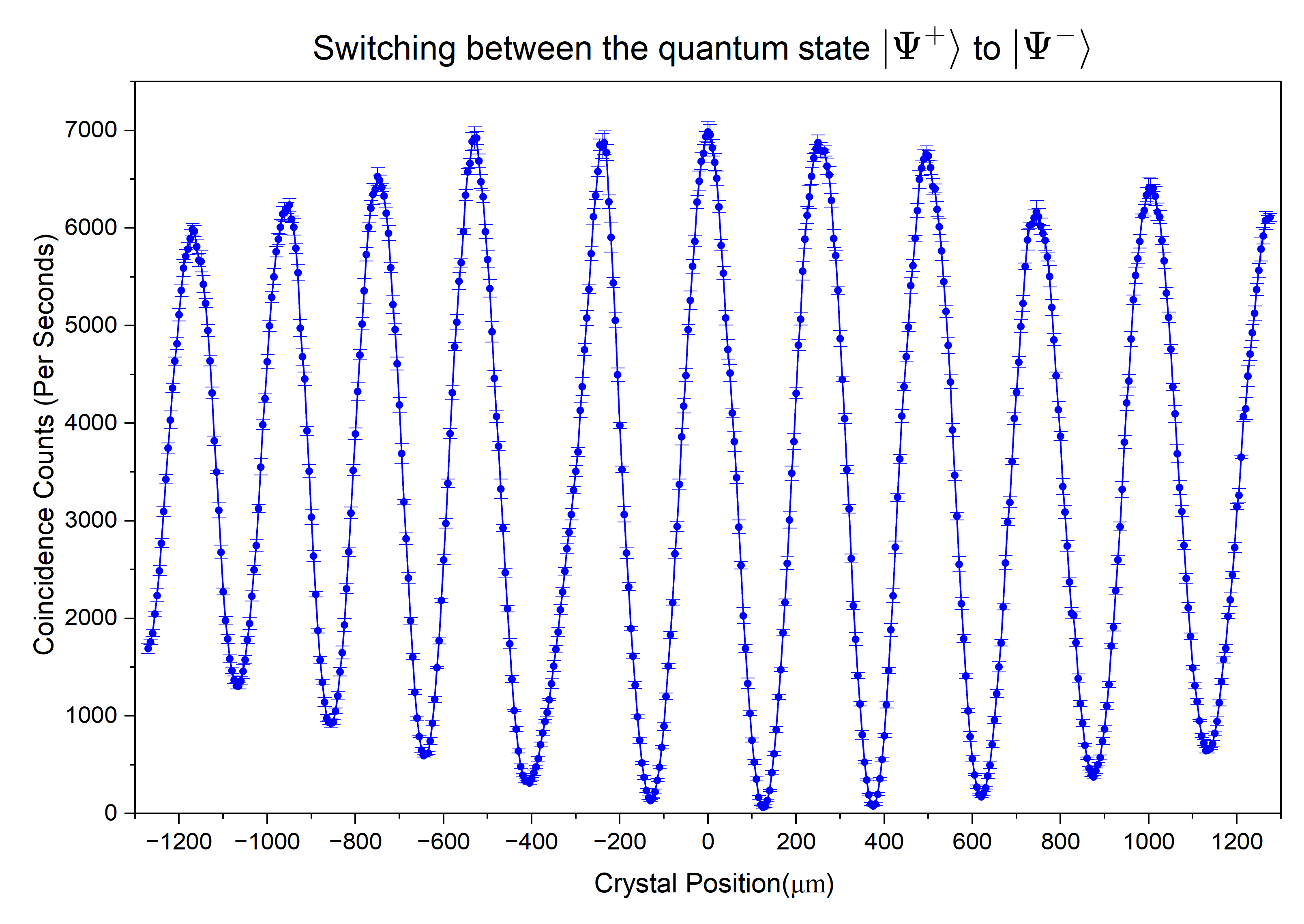}
    \caption{Projection of Bell state $|\psi\rangle$ onto the $|+_s+_i\rangle$ basis state. The balanced position of the crystal is considered to be zero.}
    \label{fig:Switching_between_the_quantum_state_Psi}
\end{figure}

Experimentally, uniform crystal translation is implemented with a linear translation stage fitted with a piezoelectric motor. The crystal translation periodically switches between the Bell states $|\phi^+\rangle$ and $|\phi^-\rangle$ (and $|\psi^+\rangle $ and $|\psi^-\rangle)$ at regular intervals. The coincidence counts for the Bell states $|\phi\rangle$ and $|\psi\rangle$ when projected onto the diagonal basis is given in Fig.~\ref{fig:Switching_between_the_quantum_state_Phi} and Fig.~\ref{fig:Switching_between_the_quantum_state_Psi}, respectively.

\begin{figure}
    \centering
    \includegraphics[width=1\linewidth]{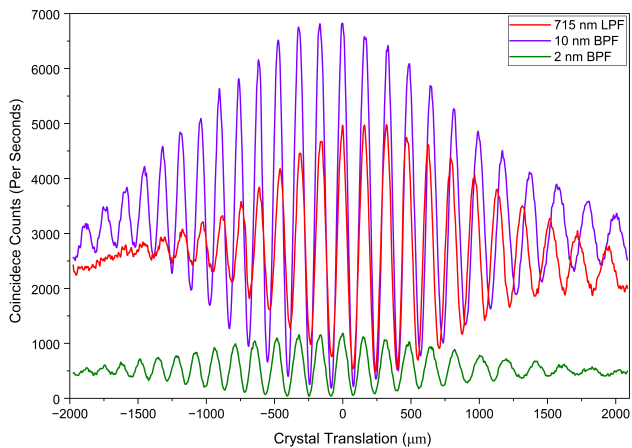}
    \caption{Effect of crystal translation on the coincidence counts while projecting the Bell states onto the $|+_s +_i \rangle$ state for $|\phi\rangle$. The coincidence counts are measured using three different spectral filters  (1) 715 nm long-pass filter, (2) 10 nm band-pass filter and (3) 2 nm band-pass filter.}
    \label{fig:3 filters diagonal projection}
\end{figure}

It is observed from the experimental results that the switching between the $|\phi^+\rangle$ ($|\psi^+\rangle$) and  $|\phi^-\rangle$ ($|\psi^-\rangle$) takes place at the regular interval of the $122 \pm 14 ~ \mu m$. In order to rule out any changes in the oscillation period due to the spectral distribution of the source we have carried out the experiment with the three different filters. Measurements are carried out by placing three different filters: 1) a longpass filter (LPF) with a cutoff wavelength at $715$ nm, 2) a bandpass filter (BPF) with $10$ nm bandwidth and 3) a bandpass filter (BPF) with $2$ nm bandwidth. We have observed no change in the oscillation period between orthogonal Bell states, and it remains the same as $122\pm14 ~ \mu m$ as we can observe in Fig. \ref{fig:3 filters diagonal projection}.

Thus, we see that the experimental curves generated by crystal translation are in qualitative agreement with theoretical predications. The reason for the observed drop in coincidence counts while moving away from the balanced position is explained later based on the observed EMCCD images. In order to carry out the complete certification of the process, Bell state measurements and QST are discussed in the following sections.

\section{ The Bell state measurement}

In the previous section, we demonstrated the effect of crystal translation from the balanced position on the coincidence count rate while projecting onto the $|+_s+_i\rangle$. In the current section, we provide the complete details of the Bell state measurement setup for the monolithic SPDC source. We shall further provide an unimpeachable certification of the states through the Bell state measurement.

\begin{figure}
    \centering
    \includegraphics[width=1\linewidth]{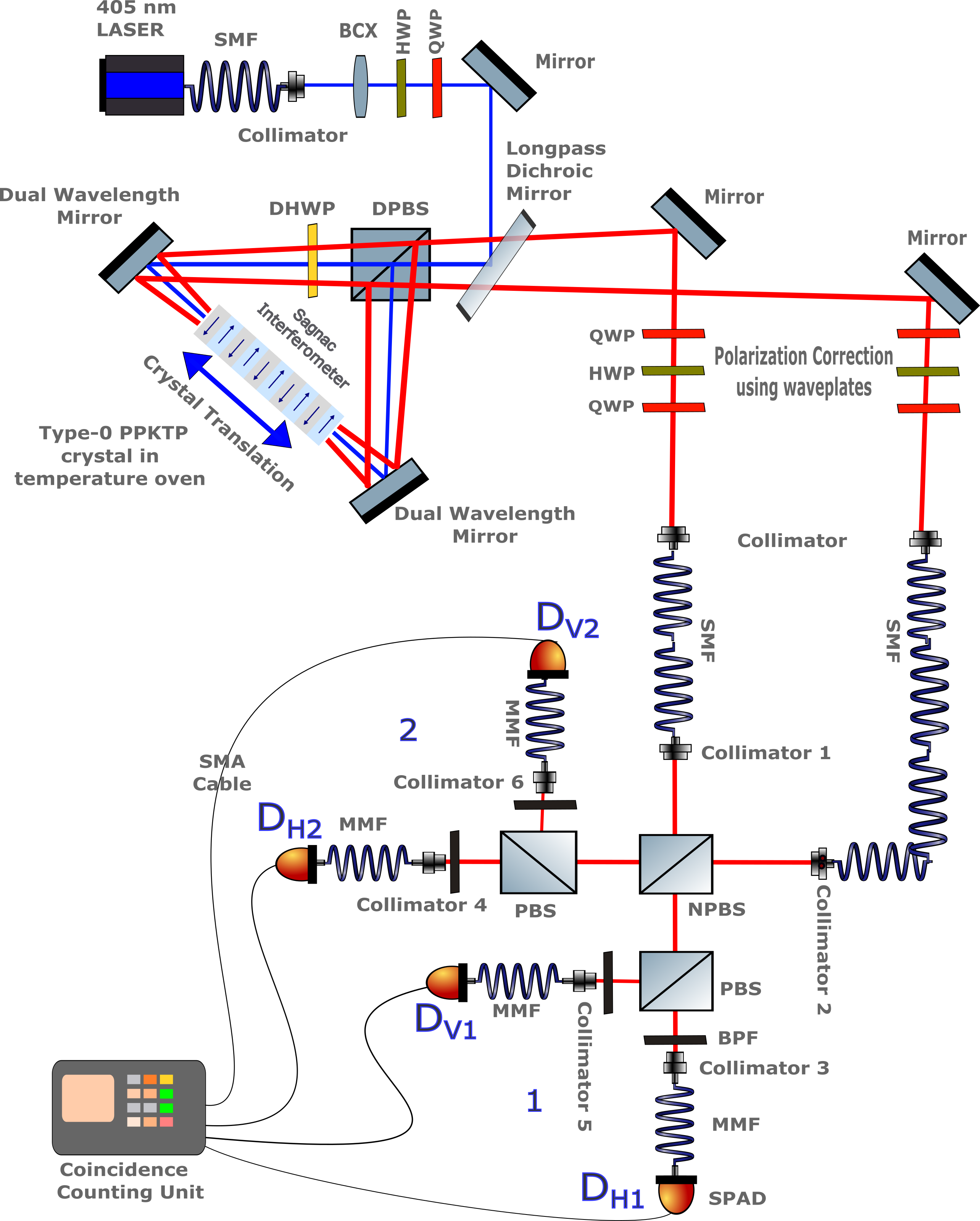}
    \caption{Schematic diagram of the Bell-state Measurement setup}
    \label{fig:BSM}
\end{figure}

The schematic of the experimental setup for the Bell state measurement is given in Fig.~\ref{fig:BSM}. The first and crucial step to perform the Bell state measurement is to achieve the Hong-Ou-Mandel interference (HOMI) between the signal and idler photons. In the HOMI setup, the signal and idler photons are allowed to interfere at the $50:50$ beam splitter. For the perfectly indistinguishable signal and idler photons, the HOM dip will be observed by balancing the path length between them. The Bell state measurement requires two additional polarizing beam splitters (PBS) kept in each arm of the output port of the Hong-Ou-Mandel interferometer. Photons exit through this output ports of the PBS are collected through multi-mode fibers and connected to SPAD units to obtain single-photon counting measurements. All four SPAD units are connected to the coincidence counter, which provides two-fold to four-fold coincidences.

Now, the signal and idler arms of the high-brightness entangled photon source are connected to the Bell state measurement setup. The EPS can generate all four Bell states ($|\phi^\pm\rangle$, $|\psi^\pm\rangle$) just by translating the crystal from its balanced position. From eqn~\ref{state}, it is clear that the quantum state $|\psi^-\rangle$ has the fermionic wavefunction, and the remaining Bell states ($|\phi^\pm\rangle$, $|\psi^+\rangle$) have the Bosonic wavefunction. If the fermionic state $|\psi^-\rangle$ is sent to the Bell state measurement setup, two photons will display the photon antibunching effects and exit through two different output ports of the beam splitter. For the fermionic state $|\psi^-\rangle$, the coincidence is always guaranteed between the detectors $D_{H_1}-D_{V_2}$ or $D_{V_1}-D_{H_2}$. This confirms the detected state is the maximally entangled Bell state $|\psi^-\rangle$. 

The bosonic wavefunction $|\psi^+\rangle$ shows the photon bunching behaviour in the Bell state measurement setup and leave the same output port of the beam splitter. For this Bell state $|\psi^+\rangle$, the coincidence is always guaranteed between the detectors $D_{H_1}-D_{V_1}$ or $D_{H_2}-D_{V_2}$. If the remaining two Bell states $|\phi^\pm\rangle$ are sent to the Bell state measurement setup, they exit through the same output port of the beam splitter. Since both  photons have the same polarization, no coincidence counts are observed.

Let us start with the maximally entangled Bell state $|\psi^+\rangle$. The crystal translation results in introducing the additional phase shift $\phi$. Now the quantum state of the system is given by,
\begin{equation}
        |\psi\rangle = \frac{1}{\sqrt{2}} \left(|H_sV_i\rangle + e^{i \phi} |V_sH_i\rangle\right)
\end{equation}

Now, continuous translation of the crystal from its balanced position results in the generation of $|\psi^+\rangle$ (whenever the phase difference is $0$ or  multiples of $2\pi$) and $|\psi^-\rangle$ (whenever the phase difference is odd multiples of $\pi$) states. This results in photon bunching for the $|\psi^+\rangle$ state and photon antibunching for the $|\psi^-\rangle$ state.

\section{Experimental setup}

A single longitudinal mode diode laser with a central wavelength $405$ nm and a spectral bandwidth $0.018$ nm is used as a pump laser. A type-0 quasi-phase matched PPKTP crystal with dimensions $2\times1\times 30$ mm and the poling period $\Lambda=3.425~\mu m$ is used for down-conversion. A half-wave plate (HWP) is used to prepare the diagonal polarization, and the quarter-wave plate (QWP) is used to correct the ellipticity in the polarization. The PPKTP crystal is kept inside the polarization Sagnac interferometer. The diagonally polarized pump photon is split into horizontal and vertical components and each component is interacting with the PPKTP from either side to produce $|H_s\rangle \otimes |H_i\rangle$ and $|V_s\rangle \otimes |V_i\rangle$ components. The downconverted $|H_s\rangle \otimes |H_i\rangle$ and $|V_s\rangle \otimes |V_i\rangle$ cones overlapped properly and emerged at the same output port of the Dual polarizing beam splitter (DPBS). Photons from two diametrically opposite points of the cone are collected through single-mode fibers (SMF). The output state of the two-qubit polarization entangled photons is given by, 
\begin{equation}
    |\phi^+\rangle = \frac{1}{\sqrt{2}} \left(|H_s\rangle \otimes |H_i\rangle + |V_s\rangle \otimes |V_i\rangle\right)
\end{equation}
One HWP is kept in the idler path. When it is oriented at $45^\circ$ angle, it produces an orthogonal polarization for the idler photon and the final polarization state of the system changes to,
\begin{equation}
    |\psi^+\rangle = \frac{1}{\sqrt{2}} \left(|H_s\rangle \otimes |V_i\rangle + |V_s\rangle \otimes |H_i\rangle\right)
\end{equation}
The PPKTP crystal is kept exactly at the centre of the interferometer. Translating the crystal in either direction in the interferometer introduces a $\phi$ phase difference between the $|HH\rangle$ and $|VV\rangle$ components. Therefore, the quantum state of the two-qubit system is given by, 
\begin{equation}
    |\phi\rangle = \frac{1}{\sqrt{2}} \left(|H_s\rangle \otimes |H_i\rangle + e^{i \phi} |V_s\rangle \otimes |V_i\rangle\right)
\end{equation}
When the crystal is moved further, the phase difference increases further and reaches $\pi$. At this stage, the state of the two-qubit system is changed to,
\begin{equation}
    |\phi^-\rangle = \frac{1}{\sqrt{2}} \left(|H_s\rangle \otimes |H_i\rangle - |V_s\rangle \otimes |V_i\rangle\right)
\end{equation}

If the half-wave plate is rotated to $45^\circ$ angle, the resultant state is, 
\begin{equation}
    |\psi^-\rangle = \frac{1}{\sqrt{2}} \left(|H_s\rangle \otimes |V_i\rangle - |V_s\rangle \otimes |H_i\rangle\right)
\end{equation}
The signal and idler photons are collected through single-mode fibers (SMF) connected to collimators. The collected signal-idler pairs are distributed over two-meter single-mode fibers for the construction of the HOMI. It is fed to a free space $50:50$ beam splitter after collimating the beam with collimators 1 and 2. The polarization state of the entangled photons undergoes unitary evolution in the SMF and its state changes during the entanglement distribution. This polarization change is corrected using the QWP-HWP-QWP combination kept in both signal and idler arms. After the polarization correction, the signal and idler photons are allowed to interfere at the $50:50$ beam splitter. A polarizing beam splitter is kept in each arm of the 50:50 beamsplitter's output port. The photons exiting through the PBS output ports are collected through a collimator and a multi-mode fiber  (MMF). The MMFs are connected to SPAD units. A bandpass filter with $10$ nm bandwidth is used in the experimental setup.  The TTL output from the SPAD units is connected with a four-channel coincidence counter to get the multi-fold coincidences.

\section{Results and Discussion}
The details of the experimental setup were provided in the previous section. As mentioned earlier, the crystal translation in the Sagnac interferometer and the half-wave plate placed in the idler arm are capable of generating all four Bell states. However, the crystal translation inside the Sagnac interferometer may affect the radius of the downconverted photon rings, due to the different distances traveled by the counterpropagating cones. This affects the overlap between the $|H_s H_i\rangle$ and $|V_s V_i\rangle$ cones at the DPBS. We provide a detailed and thorough going discussion supported by systematic measurements on the influence of the crystal translation on the overlap between the downconverted photon rings in this section. The generated maximally entangled states are verified in three stages. The first verification is performed by projecting the two-qubit systems on $|+_s +_i \rangle$ state as well as translating the crystal, and the results of it we already discussed in Section II. The second stage verification is carried out by performing the Bell state measurement. At the end, the complete quantum state tomography is performed to confirm the generated quantum state with high fidelity. 

\begin{figure*}[t]
\centering
\setlength{\tabcolsep}{2pt}

\begin{tabular}{c c c c c c}

 &  \textbf{$\bm {0~\mu m}$} & \textbf{$\bm {610~\mu m}$}  & \textbf{$\bm {1220~\mu m}$}  & \textbf{$\bm {10 \times 10^{3} ~\mu m}$}  & \textbf{$\bm {15 \times 10^{3}~\mu m}$} \\

\raisebox{1.5cm}{\textbf{H - cone}}
& \includegraphics[width=0.16\textwidth]{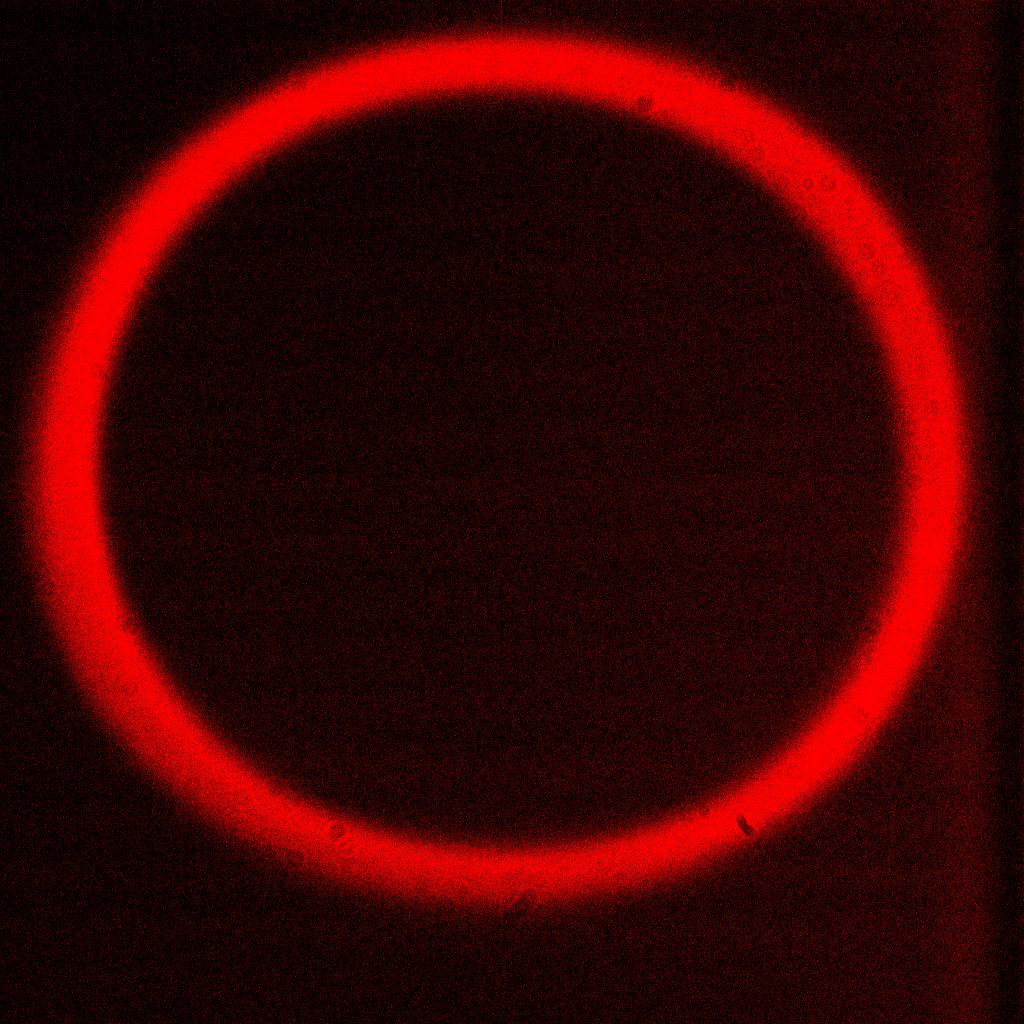} &
\includegraphics[width=0.16\textwidth]{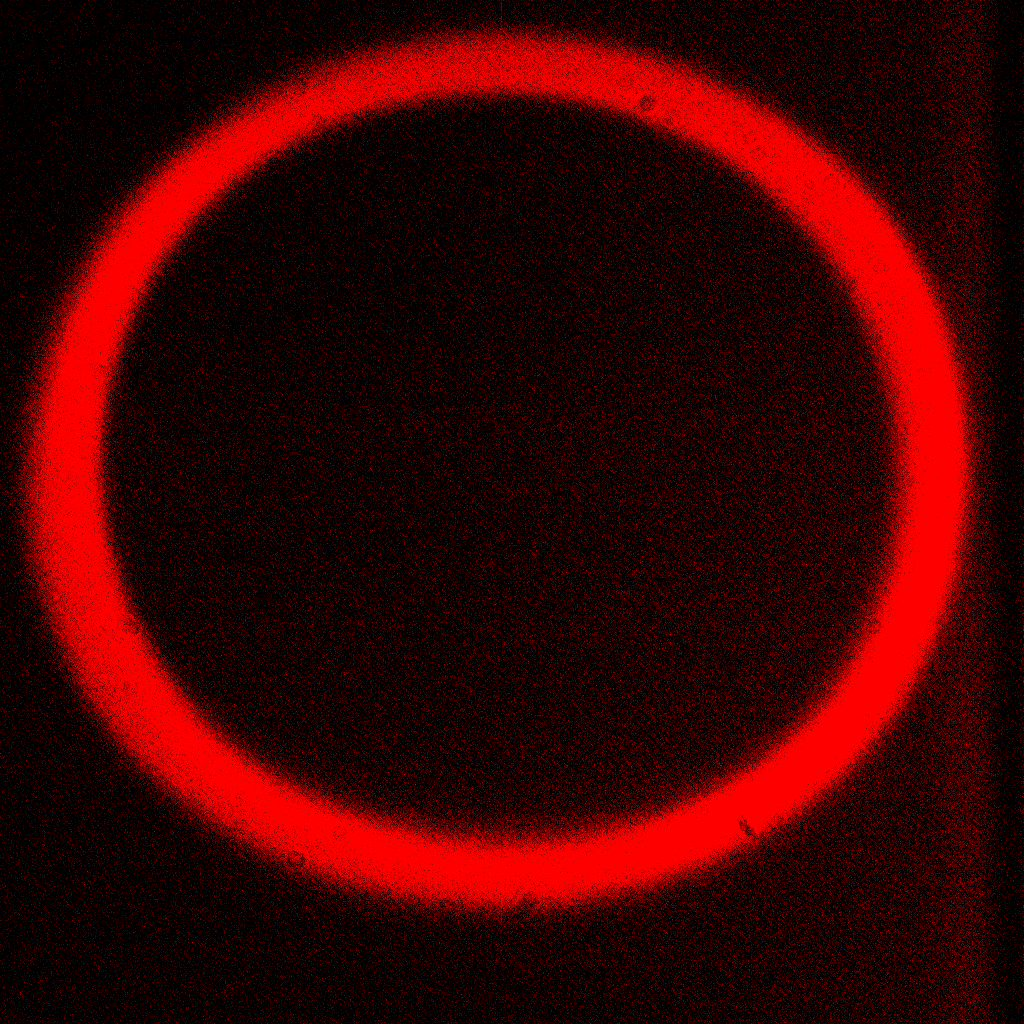} &

\includegraphics[width=0.16\textwidth]{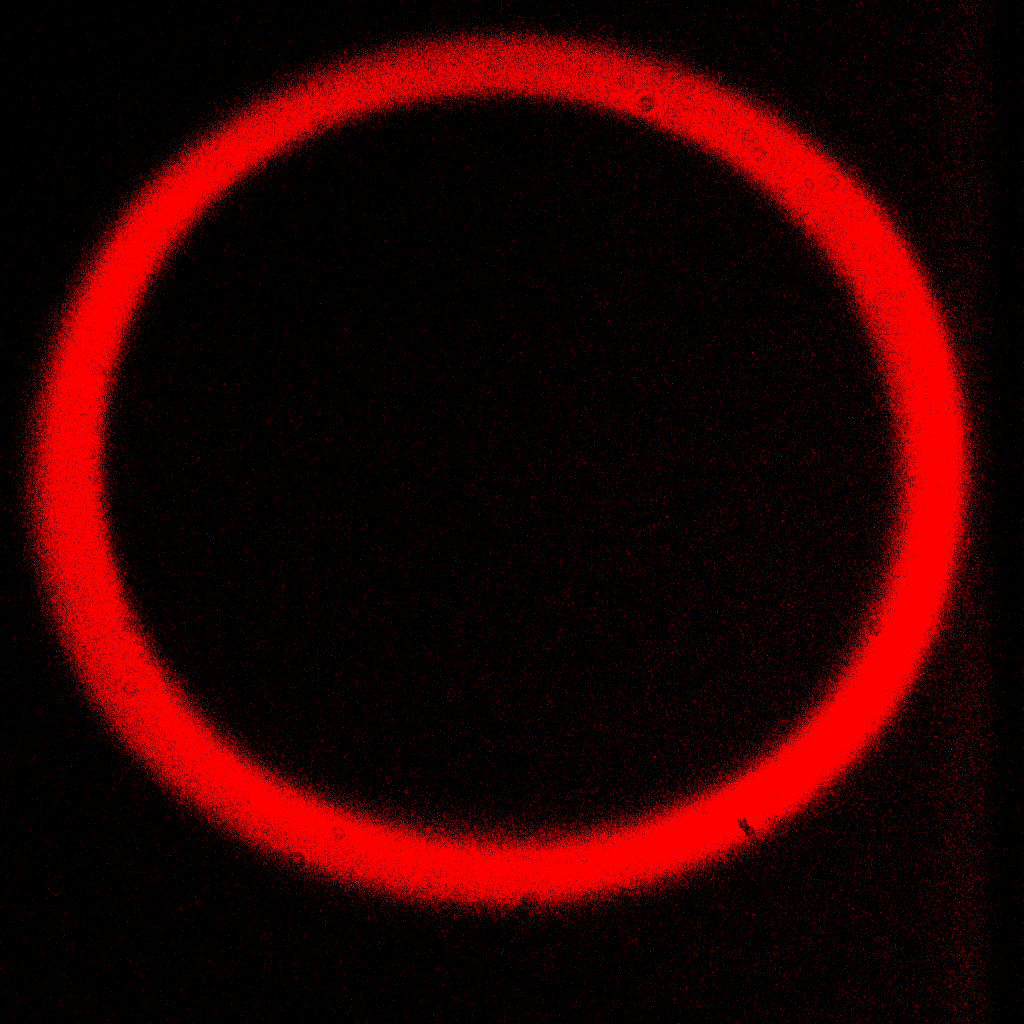} &

\includegraphics[width=0.16\textwidth]{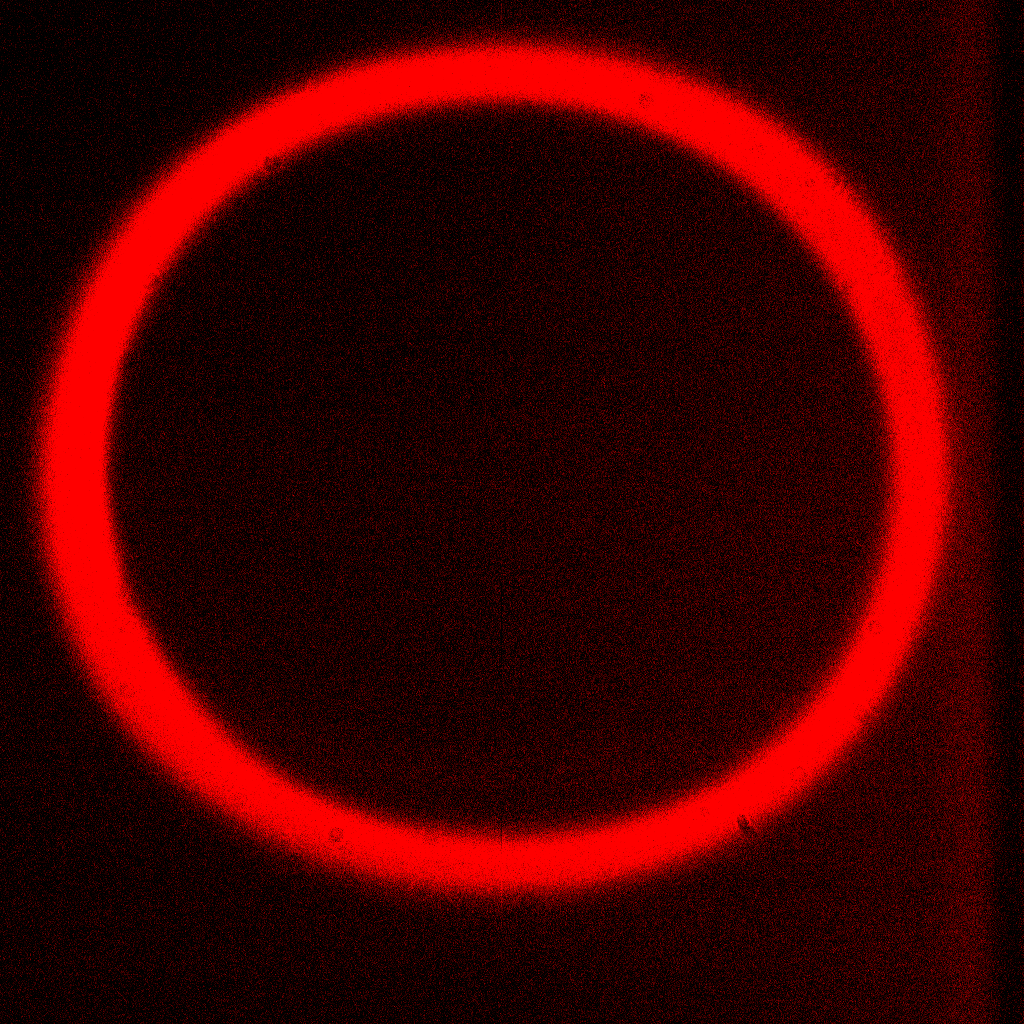} &

\includegraphics[width=0.16\textwidth]{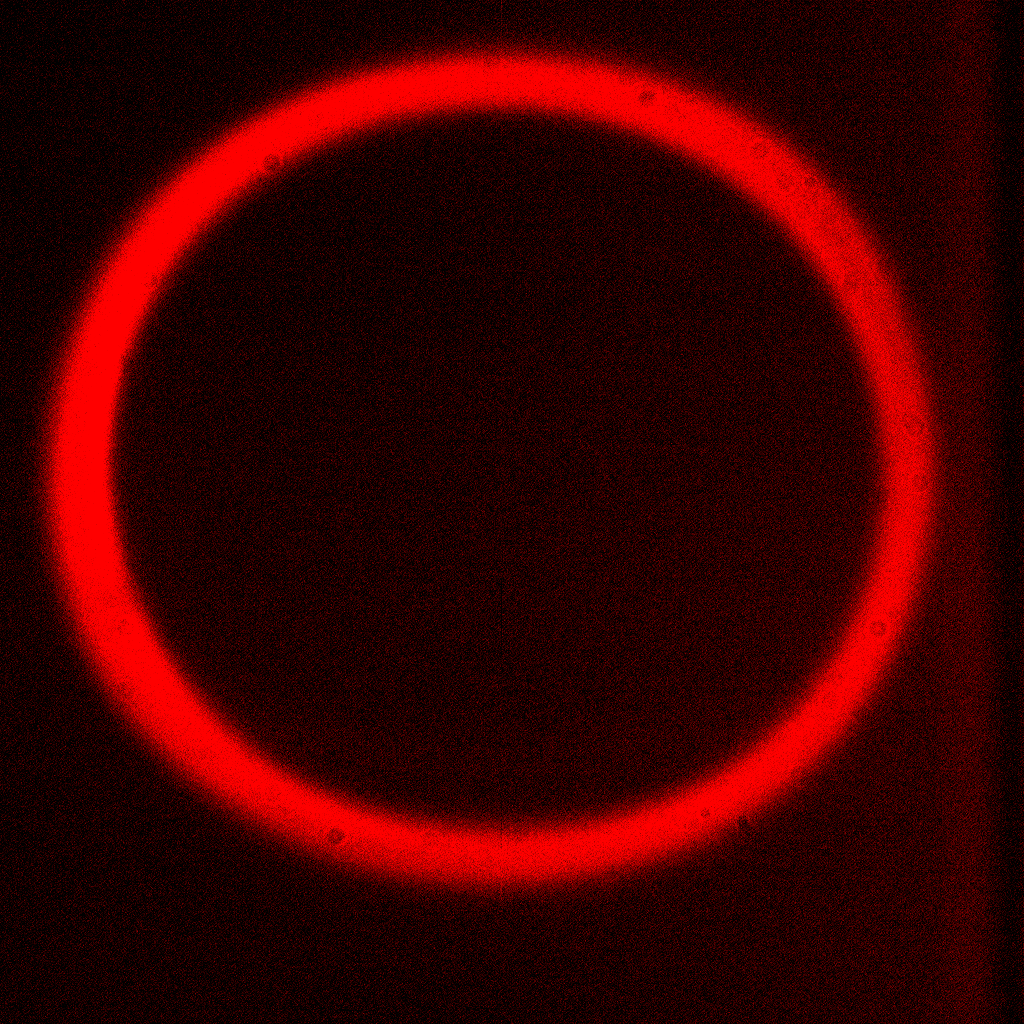} \\

%%%%%%%%%%%%%%%%%%%%%%%%
\raisebox{1.5cm}{\textbf{V - cone}} & \includegraphics[width=0.16\textwidth]{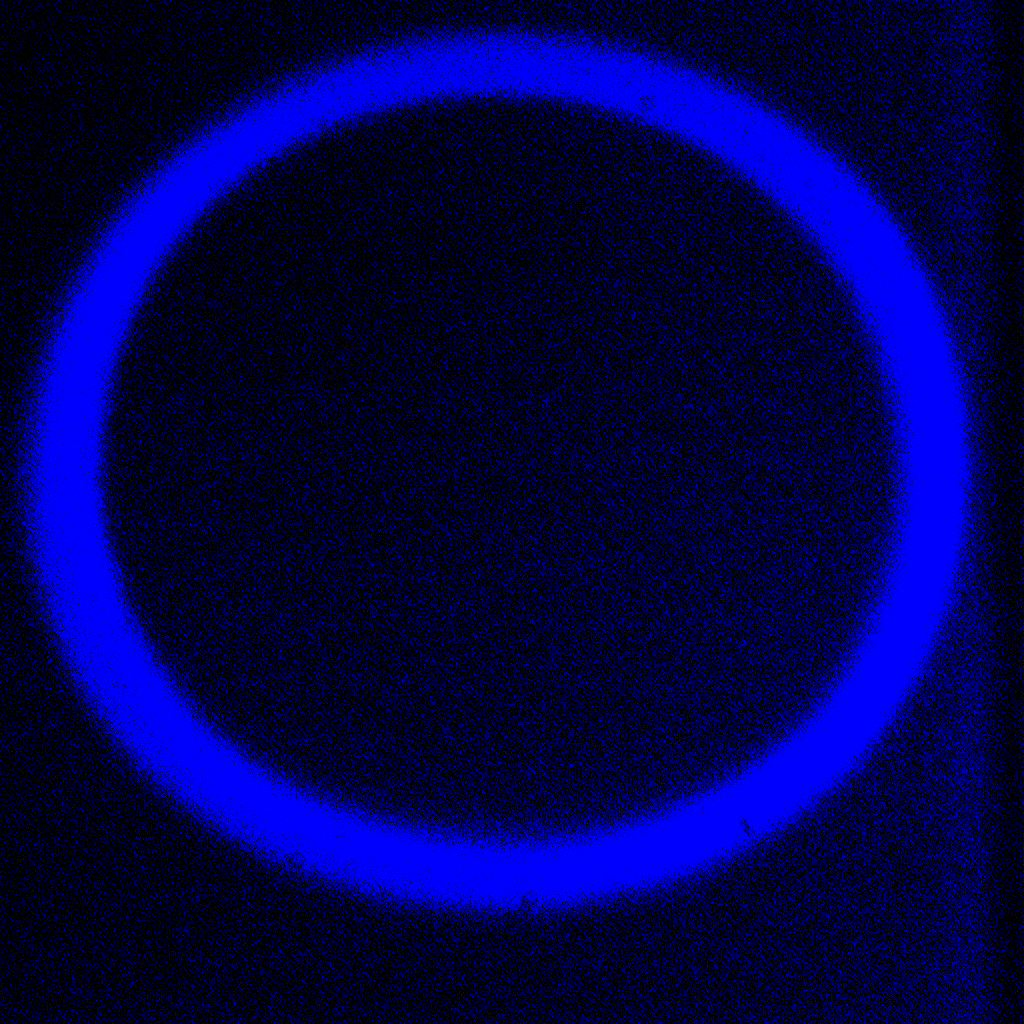} &

\includegraphics[width=0.16\textwidth]{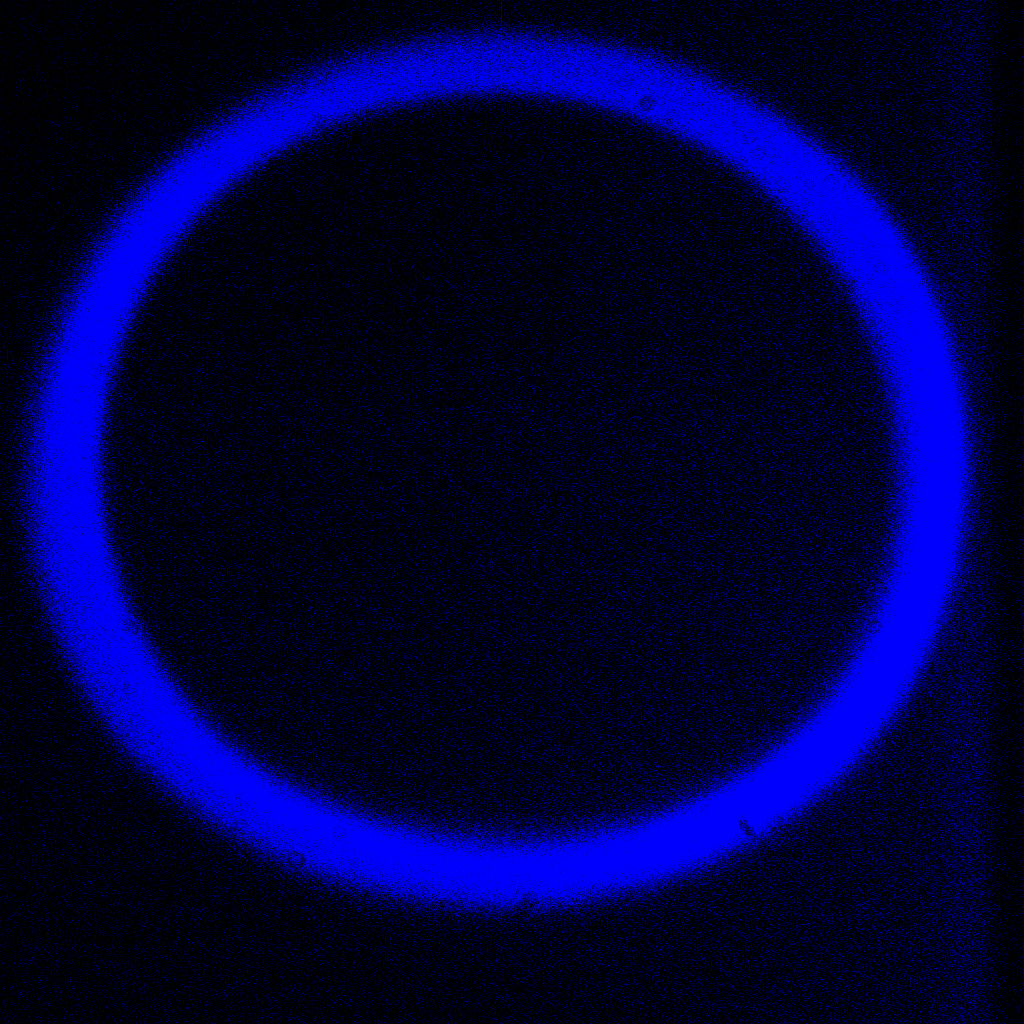} &

\includegraphics[width=0.16\textwidth]{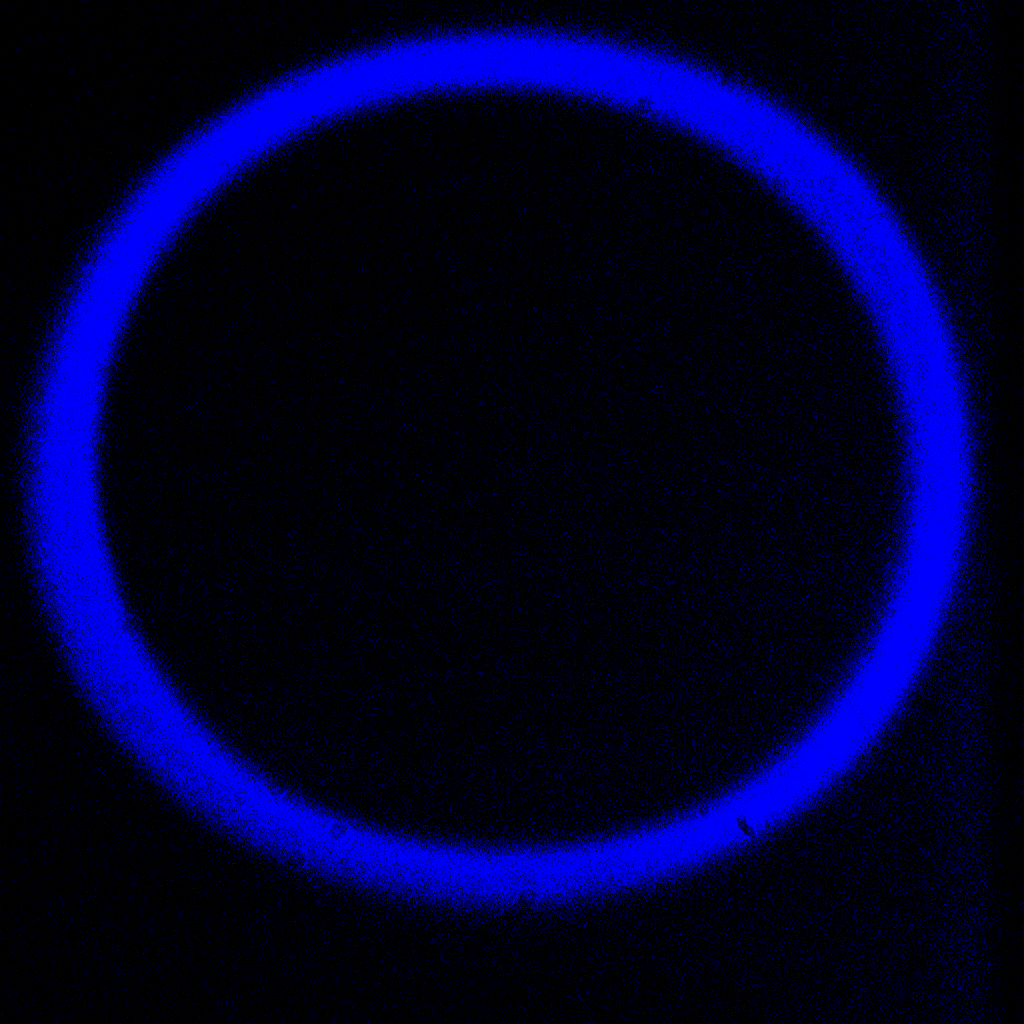} &

\includegraphics[width=0.16\textwidth]{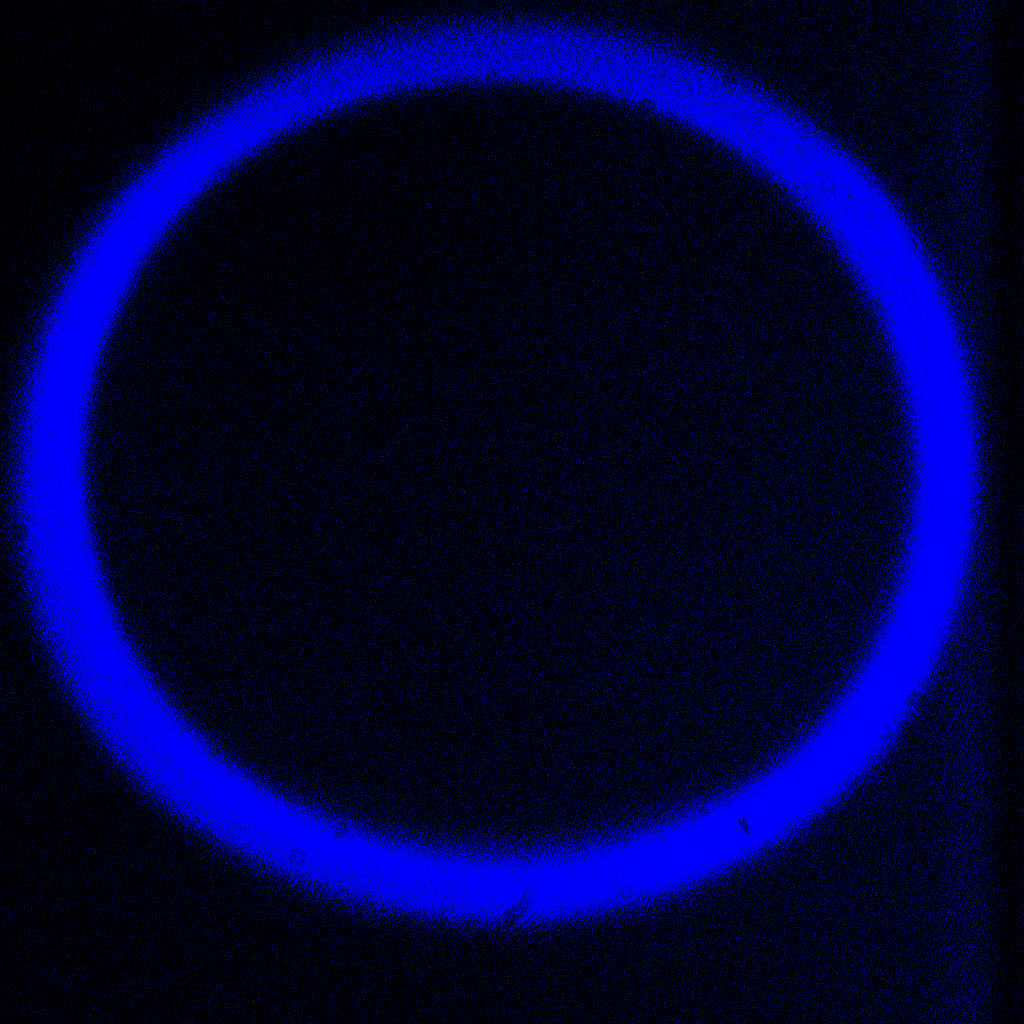} &

\includegraphics[width=0.16\textwidth]{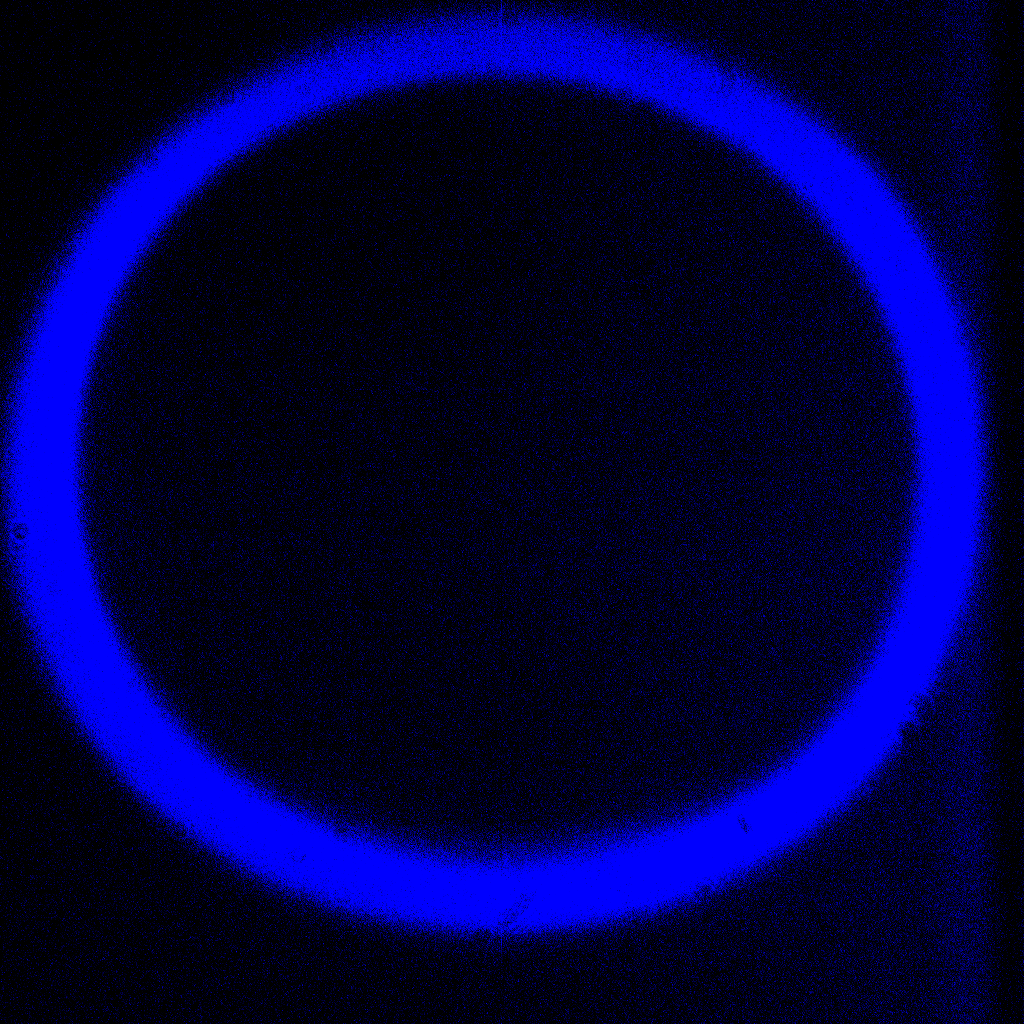} \\

    %%%%%%%%%%%%%%%%%%%%%%%%%%%%%

\raisebox{1.5cm}{\textbf{H and V - cone}} & \includegraphics[width=0.16\textwidth]{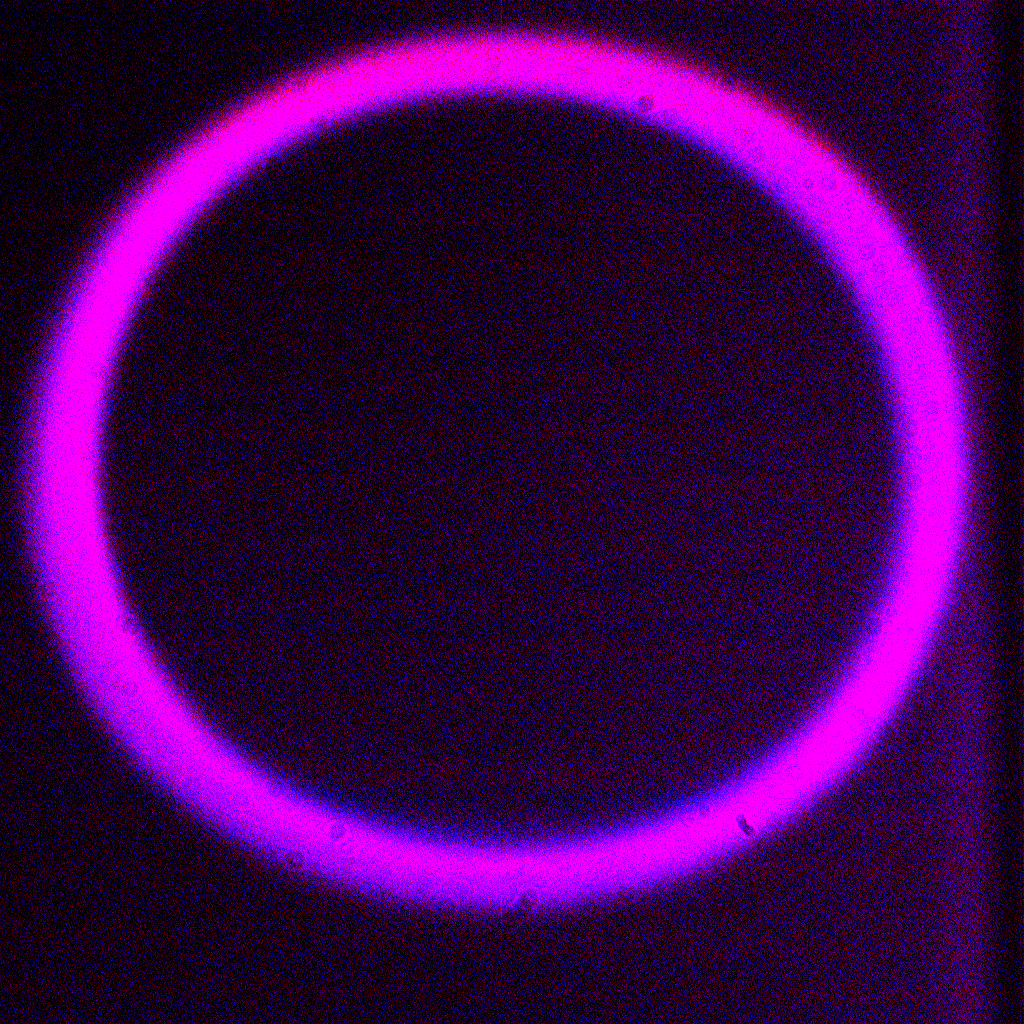} &

\includegraphics[width=0.16\textwidth]{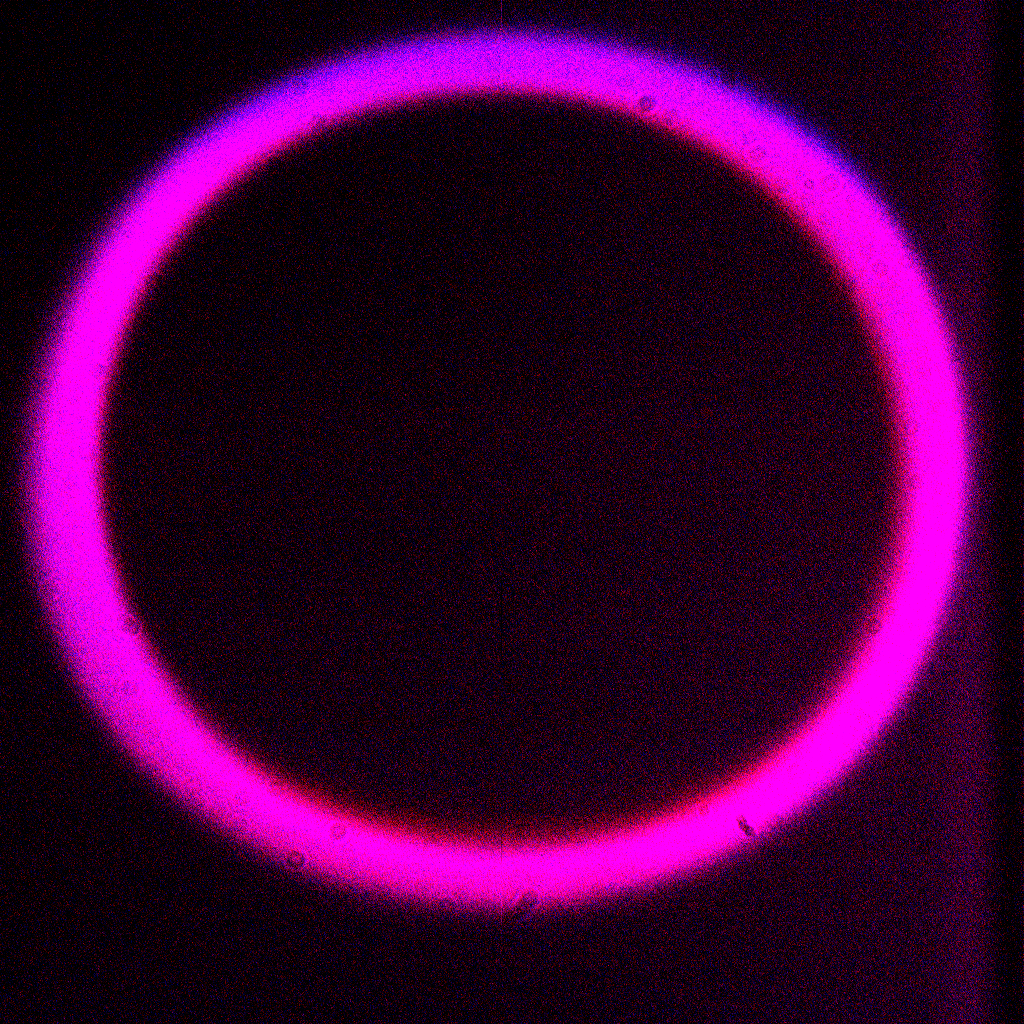} &
\
\includegraphics[width=0.16\textwidth]{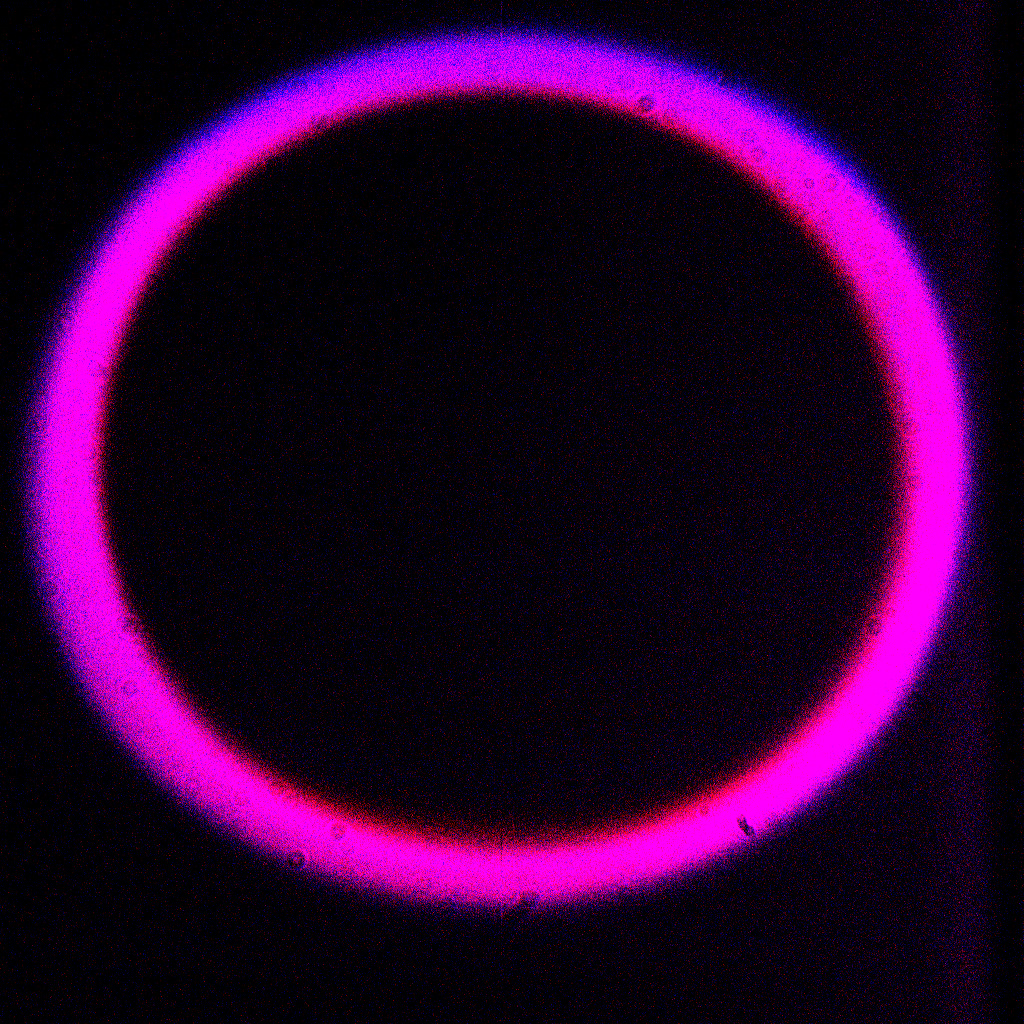} &

\includegraphics[width=0.16\textwidth]{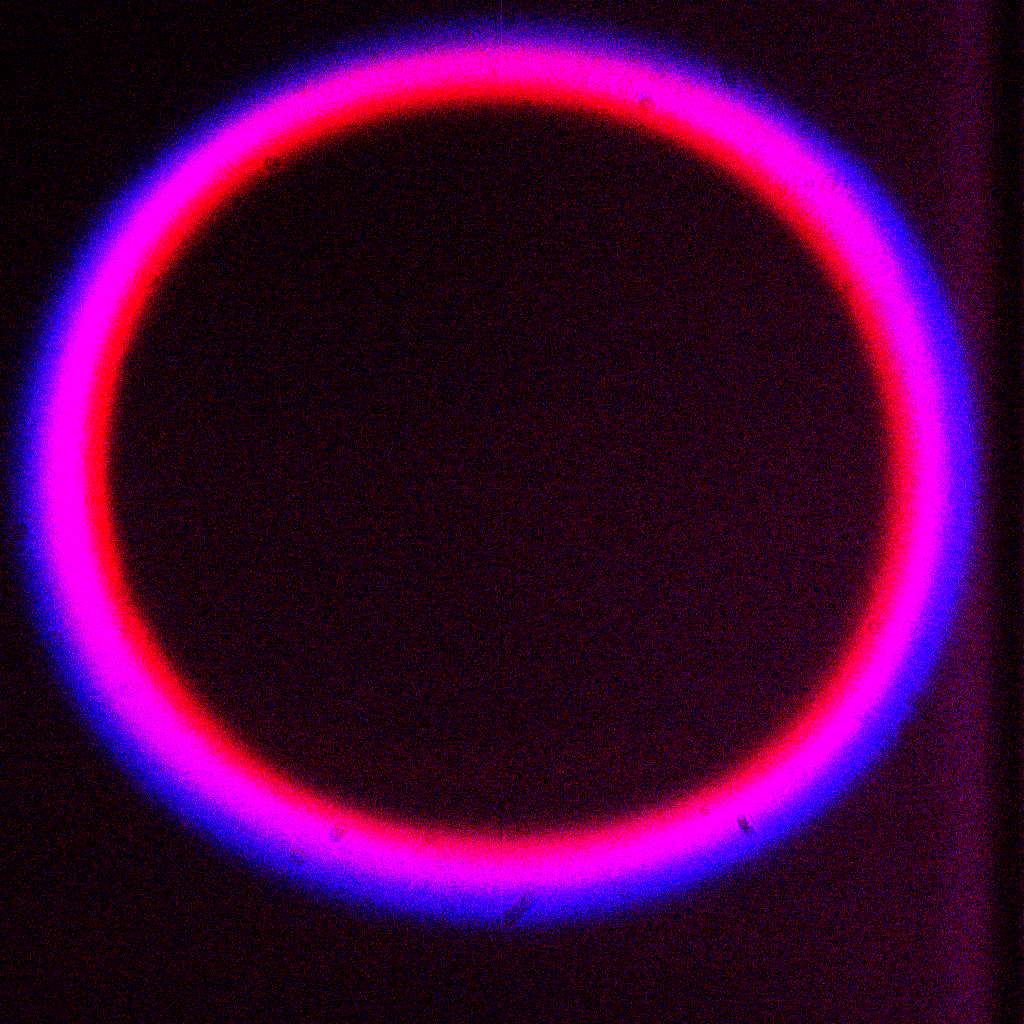} &

\includegraphics[width=0.16\textwidth]{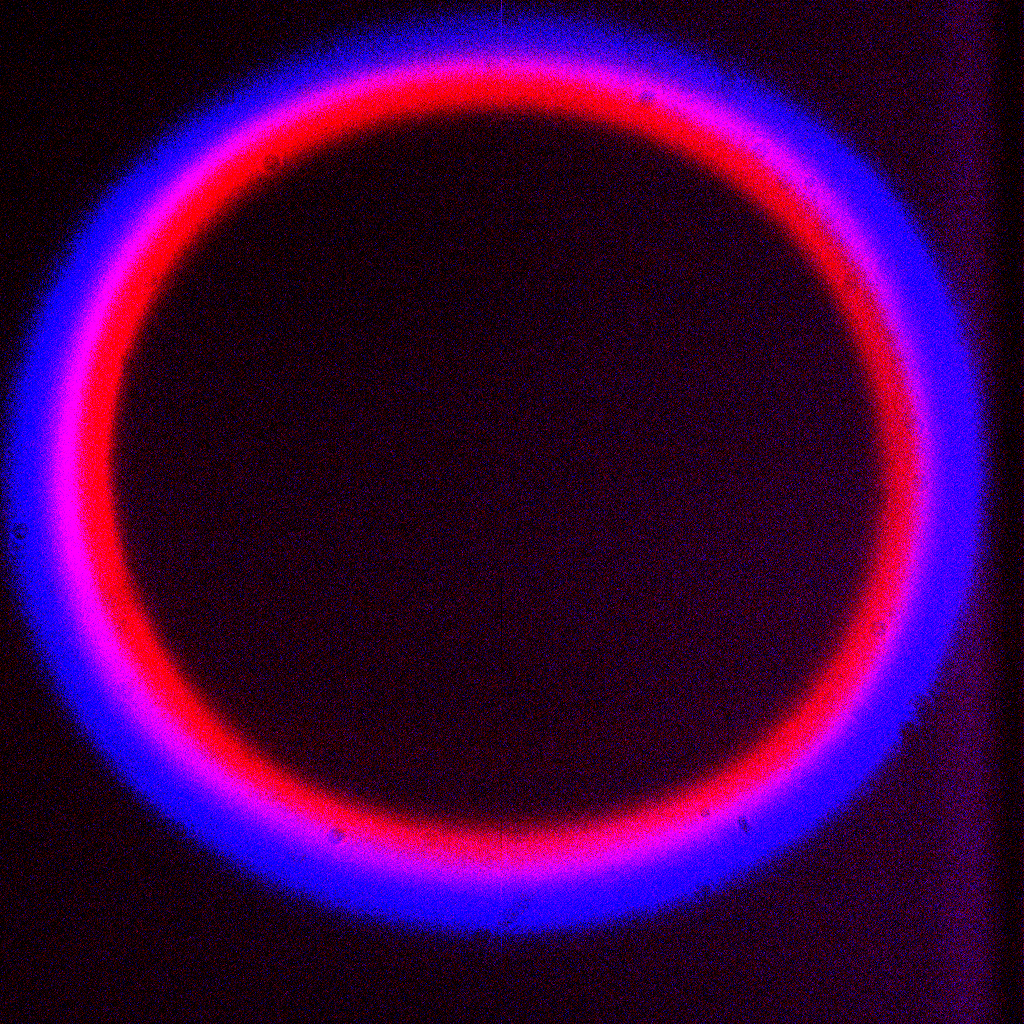} \\
%%%%%%%%%%%%%%%%%%%%%%%%%%%%%%%%%%

&  \textbf{(a)} & \textbf{(b)}  & \textbf{(c)}  & \textbf{(d)}  & \textbf{(e)} \\

\end{tabular}

\caption{Matrix of EMCCD images for different polarization settings and crystal positions from the balanced position of the Sagnac interferometer. Each row shows EMCCD images of the H - cone, V - cone and the overlapped H and V - cones, respectively. Each column corresponds to the EMCCD image taken from a different crystal translation position from the balanced position of the Sagnac interferometer. The EMCCD images from the first three columns (a - c) show that the two counter-propagating cones overlap well even after translating the crystal by $1220~\mu m$. The noticeable change in the overlap is observed at $10~mm$ and $15~mm$ as shown in column (d) and (e).}
\label{fig:EMCCD images}
\end{figure*}

\subsection{Effect of crystal translation on the radius of the downconverted photon rings}

%%%%%%%%%%%%%%%%%%%%%%%

In the Sagnac interferometer, downconverted photons are generated as a cone on either side of the PPKTP crystal. The $|H_s H_i\rangle$ cone travels in the clockwise direction, and the $|V_s V_i\rangle$ cone travels in the counterclockwise direction in the Sagnac interferometer. The transmitted $|H_s H_i\rangle$ cone and the reflected $|V_s V_i\rangle$ cones overlap at the DPBS. When the crystal is kept at the balanced position of the interferometer, there is a perfect overlap between them. The crystal translation from its central position and placing a HWP at $0^\circ$ and $45^\circ$ in the idler arm results in the generation of all four Bell states. However, the crystal translation from its balanced position in either direction increases the radius of a cone that travels a longer distance and decreases the radius of the other cone that travels a shorter distance. For example, when the crystal is translated from its balanced position in the clockwise direction, the $|V_s V_i\rangle$ cone has to travel a greater distance compared to the $|H_s H_i\rangle$ cone. Due to this, the ring radius of the $|V_s V_i\rangle$ cone increases, and the ring radius of the $|H_s H_i\rangle$ cone decreases, as given in the image taken from the EMCCD camera. 

%%%%%%%%%%%%%%%%%%%
\begin{figure*}[t]
\centering

\begin{tabular}{@{}cc@{}}

\begin{tabular}{c}
\includegraphics[width=0.495\textwidth]{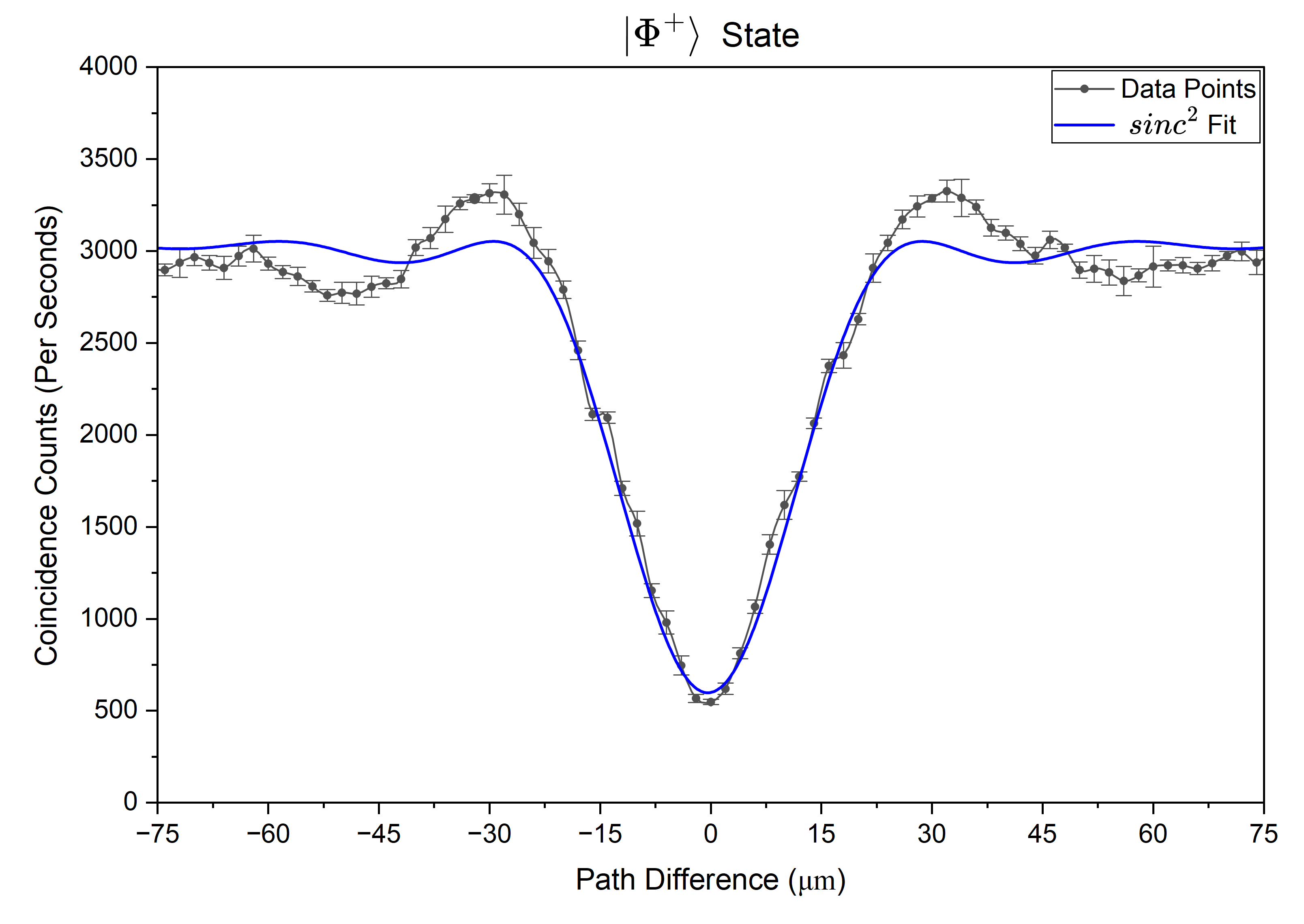} \\
(a)
\end{tabular}
&
\begin{tabular}{c}
\includegraphics[width=0.495\textwidth]{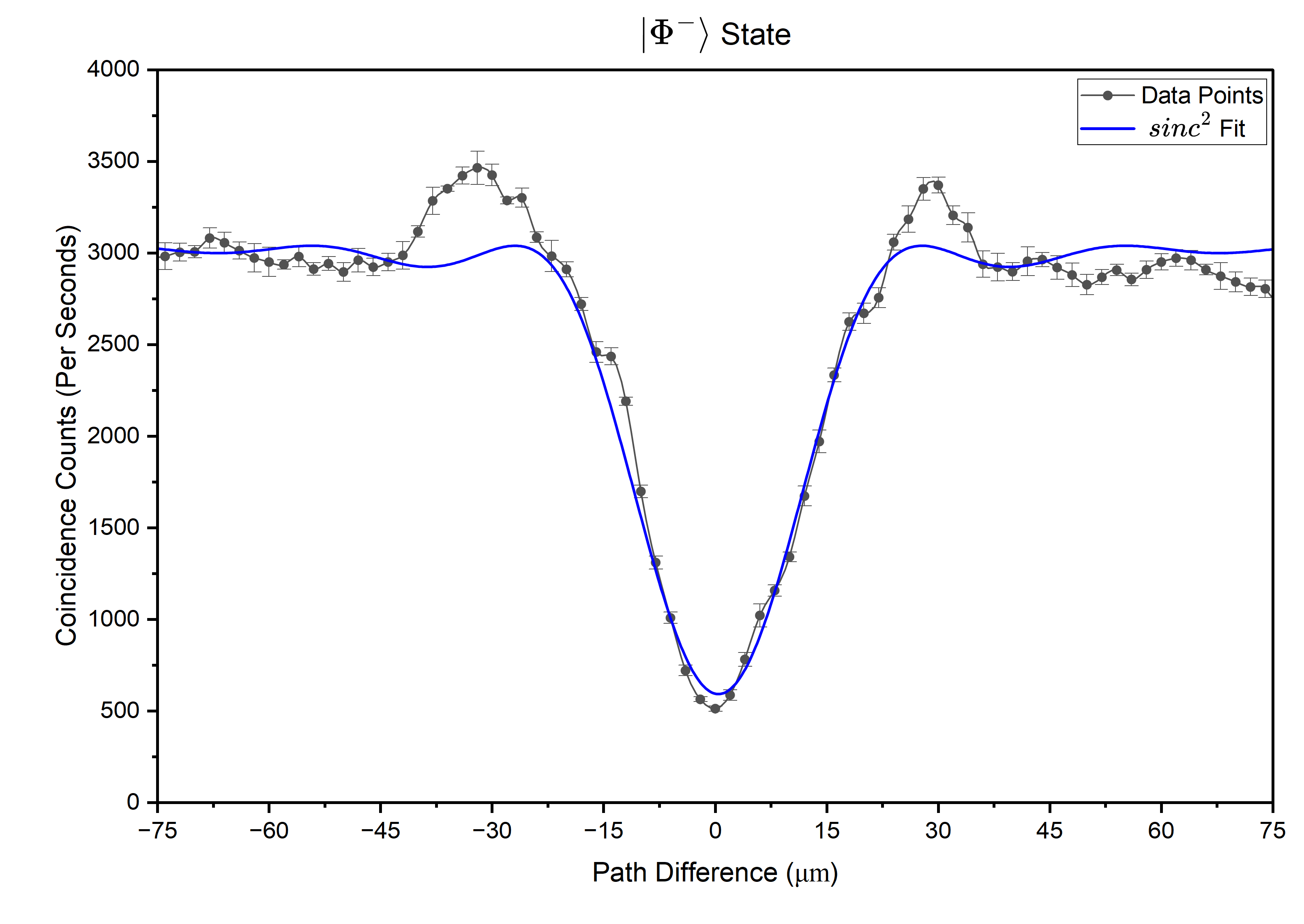} \\
(b)
\end{tabular}

\\[1.5mm]

\begin{tabular}{c}
\includegraphics[width=0.495\textwidth]{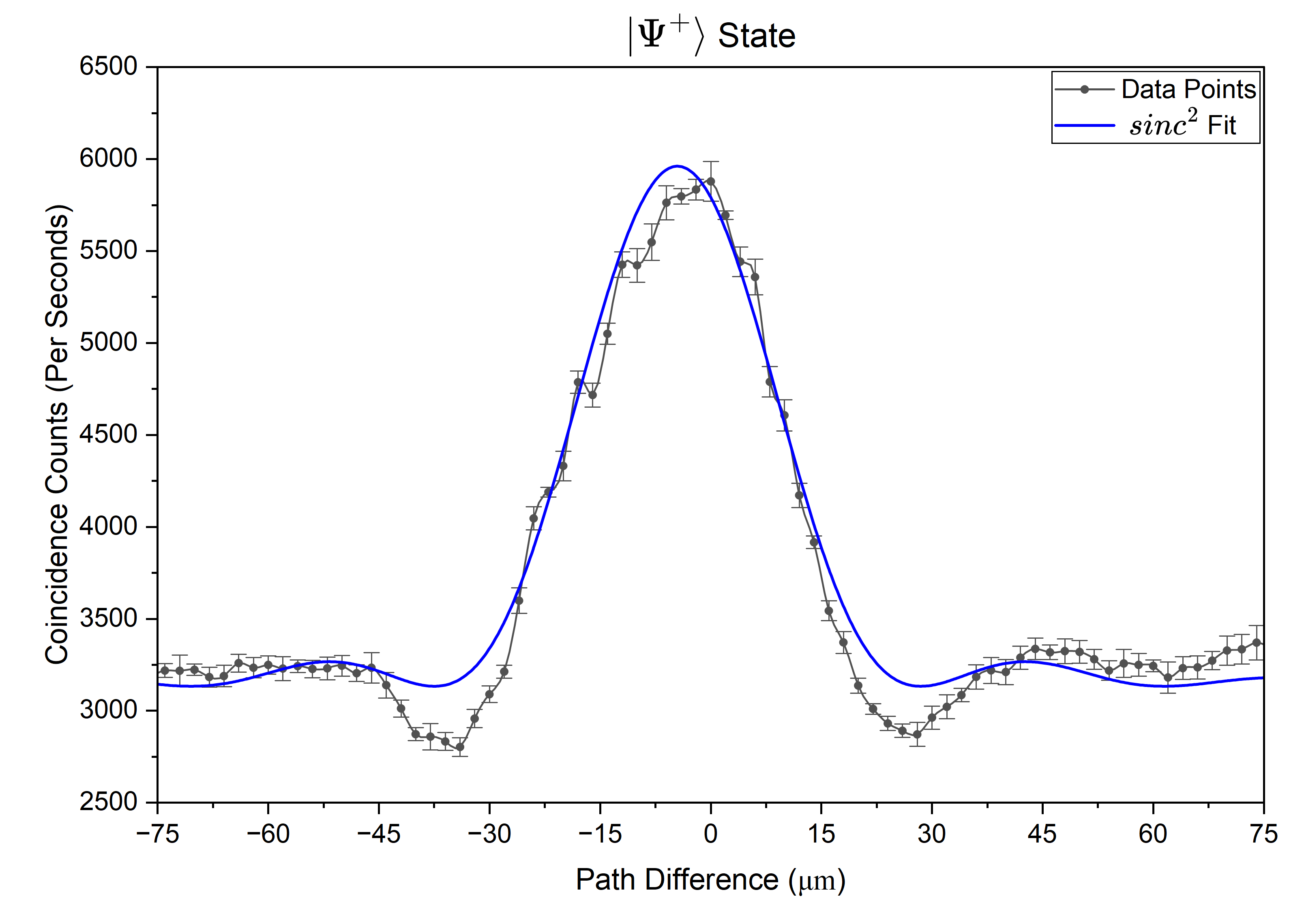} \\
(c)
\end{tabular}
&
\begin{tabular}{c}
\includegraphics[width=0.495\textwidth]{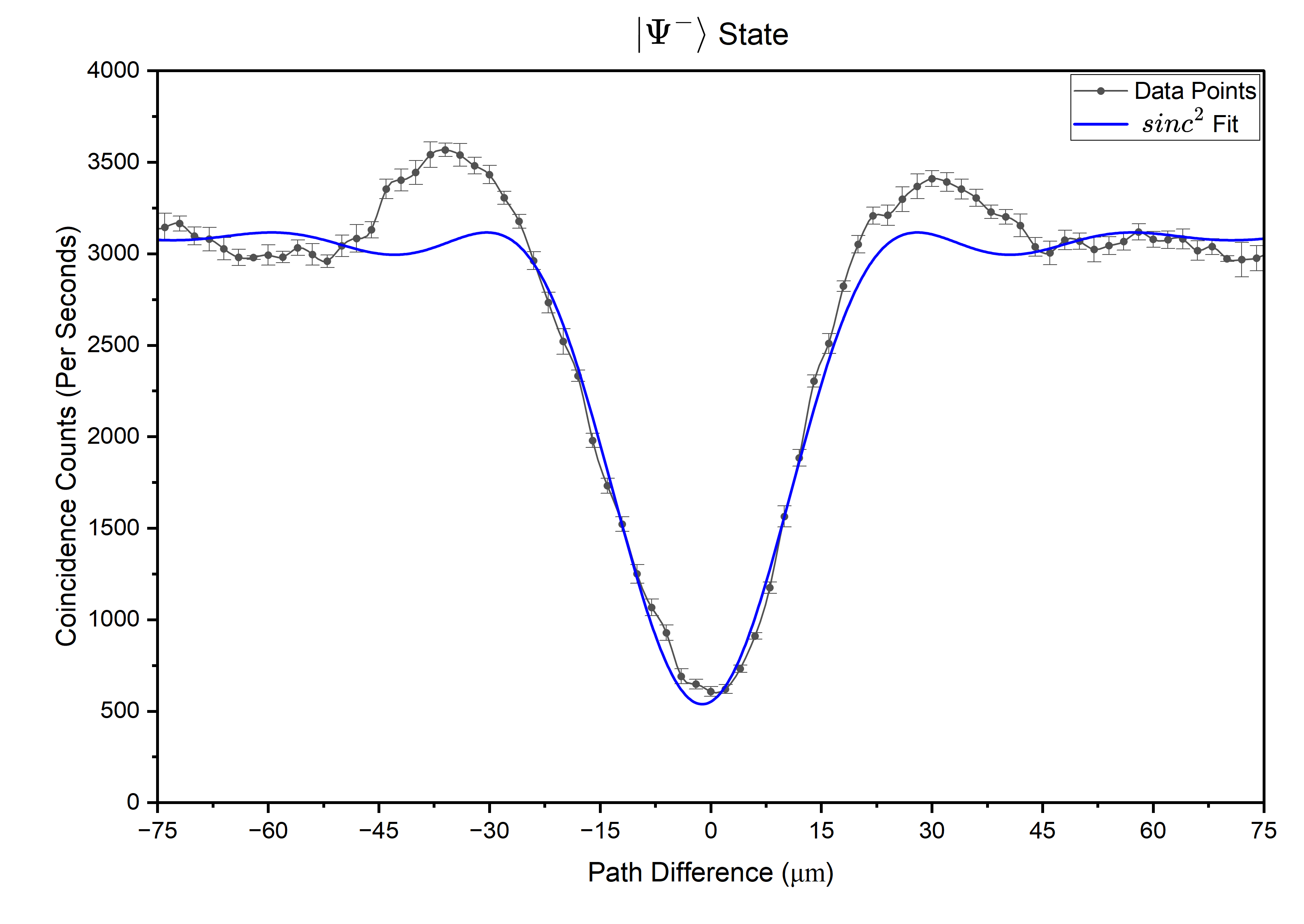} \\
(d)
\end{tabular}

\end{tabular}

\caption{
Hong-Ou-Mandel Interference (HOM) using a dielectric 50:50 beam splitter for different Bell states:
(a) $|\phi^+\rangle$ 
(b) $|\phi^-\rangle$ 
(c) $|\psi^+\rangle$ and
(d) $|\psi^-\rangle$
}
\label{fig:HOMI_2x2_full}
\end{figure*}
%%%%%%%%%%%%%%%%%%%%%%%%%%%

\begin{figure}
  
    \includegraphics[width=1\linewidth]{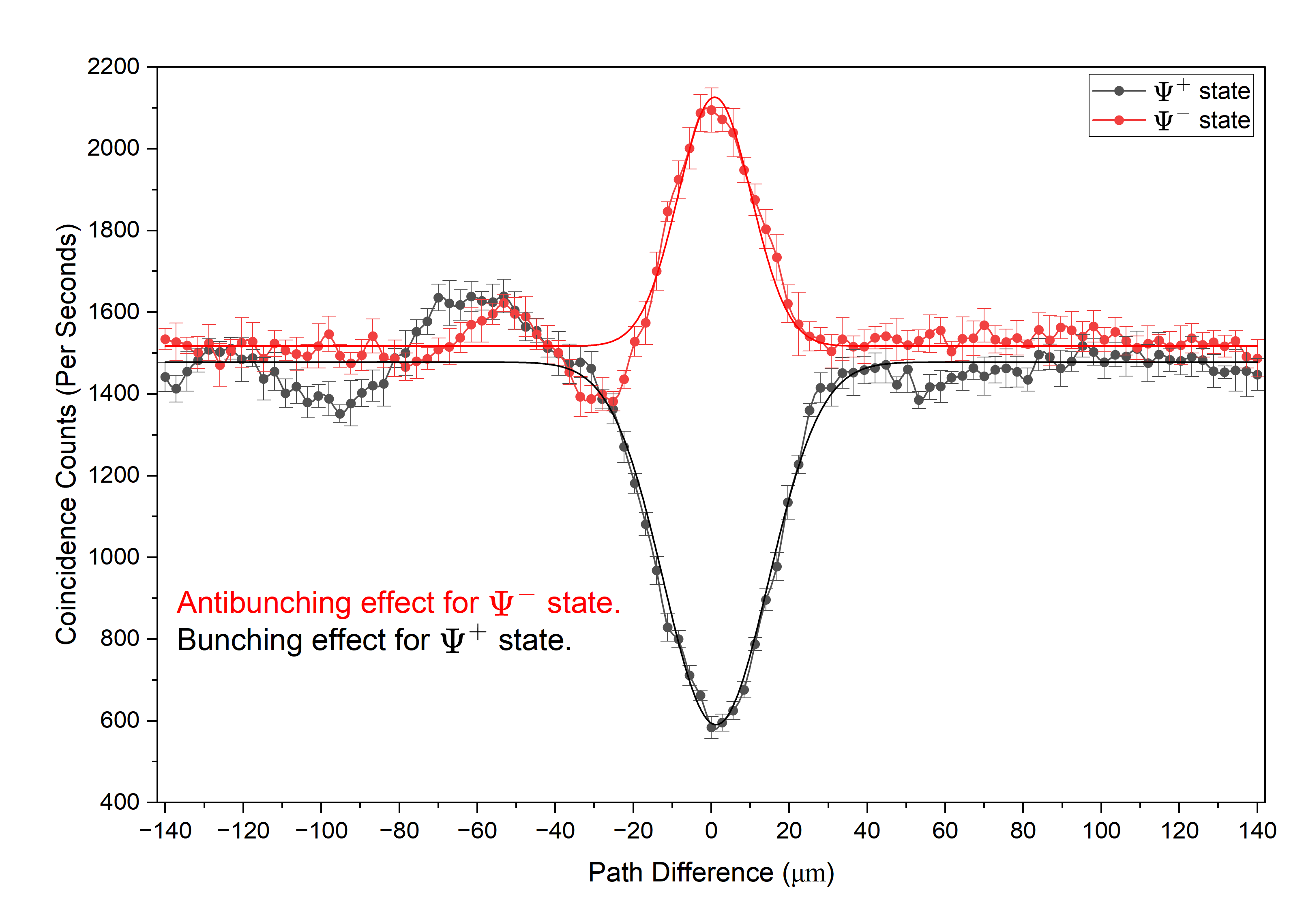}
    \caption{Hong-Ou-Mandel (HOMI) interference measured using polarization-maintaining fiber beam splitter for the Bell states $|\psi^+\rangle$ and $|\psi^-\rangle$}
    \label{fig:qutool's FBS HOM}
\end{figure}
%%%%%%%%%%%%%%%%%%%%%

The crystal is kept inside a temperature oven at $26.5^\circ$. At this temperature, the signal-idler pairs emerge at $\theta_s=-\theta_i=2.2^\circ$ with respect to the pump beam. For this angle of downconversion, the $tan~2.2^\circ = 0.03841$. The downconverted photons are collected at $15$ cm distance from the center of the crystal. The radius of the ring of the down-converted photons after $15$ cm is $5.7624$ mm. If the crystal is translated in clockwise direction by $122~\mu m$ to generate the orthogonal Bell states, the ring radius of the $|H_s H_i\rangle$ and $|V_s V_i\rangle$ cones changes to $5.7577$ mm and $5.7671$ mm, respectively. Due to the crystal translation by $122~\mu m$, the ring radius of both cones changes and the cones are moved from their central position by $4.6868~\mu m$. The ratio of the change in the ring radius to the crystal movement is insignificant. Therefore, even though the crystal is moved, the overlap between the downconverted photons is still there. This allows us to make a stable entangled photon source even though the crystal is moved to a larger distance from its balanced position. Although the crystal is translated by $1220~\mu m$, the cones are shifted from their central position only by $46.8677~\mu m$.

Figure \ref{fig:EMCCD images}, provides an EMCCD images of the down-converted photons for different crystal translation positions and polarization settings. From the EMCCD images, it is clear that even though the crystal is translated in the clockwise direction a few thousand micrometers from its central position, the overlap between the $|H_s H_i\rangle$ and $|V_s V_i\rangle$ cones is still there. This results in a negligible reduction in single and coincidence counts, but it is not completely lost. Within $1220~\mu m$ distance, the quantum state periodically switches between $|\phi^+\rangle$ and $|\phi^-\rangle$ nearly 10 times (Refer Fig.~\ref{fig:Switching_between_the_quantum_state_Phi} and Fig.~\ref{fig:Switching_between_the_quantum_state_Psi}). As given in Fig. \ref{fig:EMCCD images}, appreciable change in the overlap between the H and V cones is observed at $10~mm$ and $15~mm$. In the next section, we discuss the certification of the generated Bell state using BSM.

\subsection{Influence of the Crystal translation on the generated quantum states and Bell state measurement}

%%%%%%%%%%%%%%%%
\begin{figure*}[t]
\centering

\begin{tabular}{@{}cc@{}}

\begin{tabular}{c}
\includegraphics[width=0.48\textwidth]{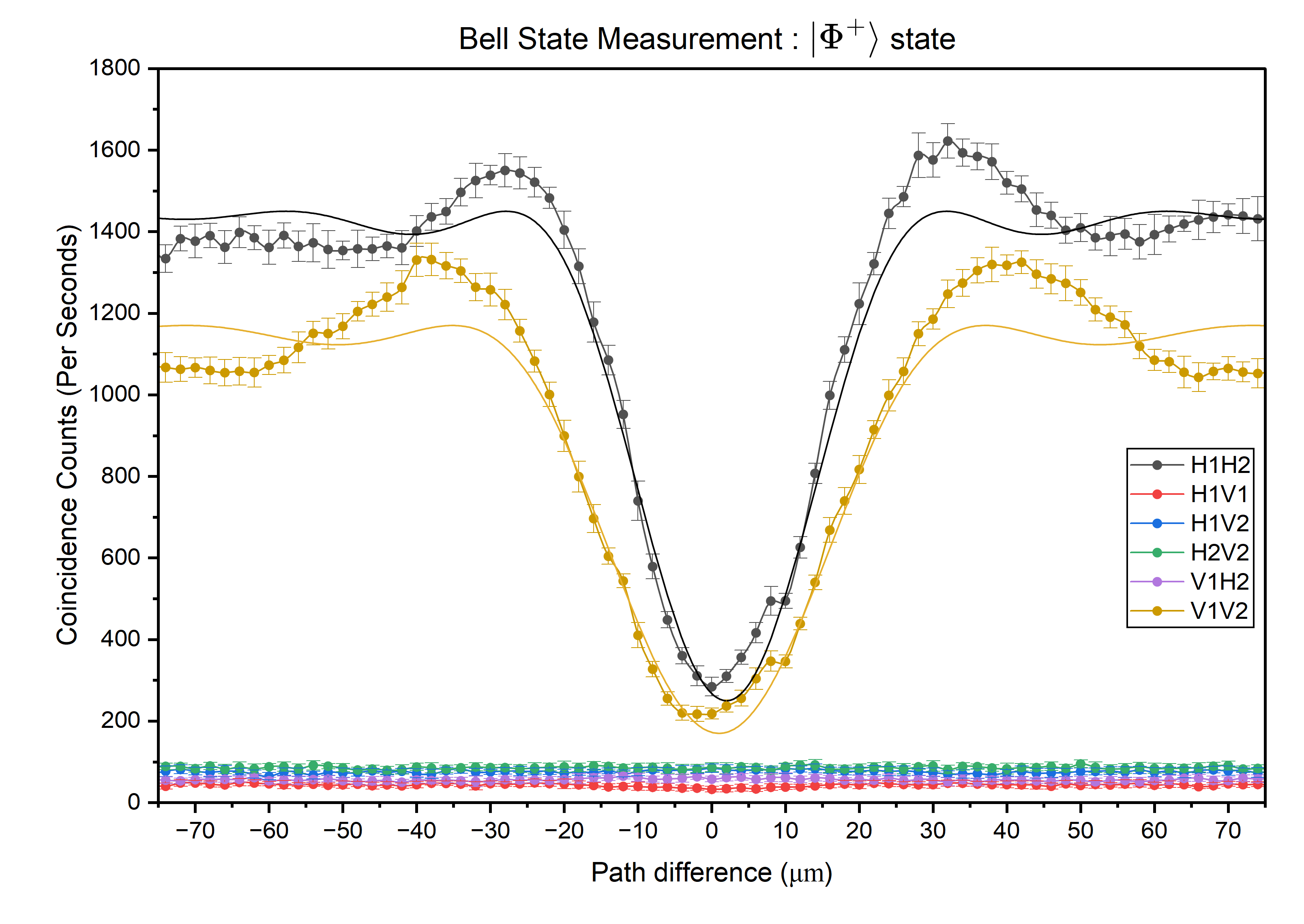} \\
(a)
\end{tabular}
&
\begin{tabular}{c}
\includegraphics[width=0.48\textwidth]{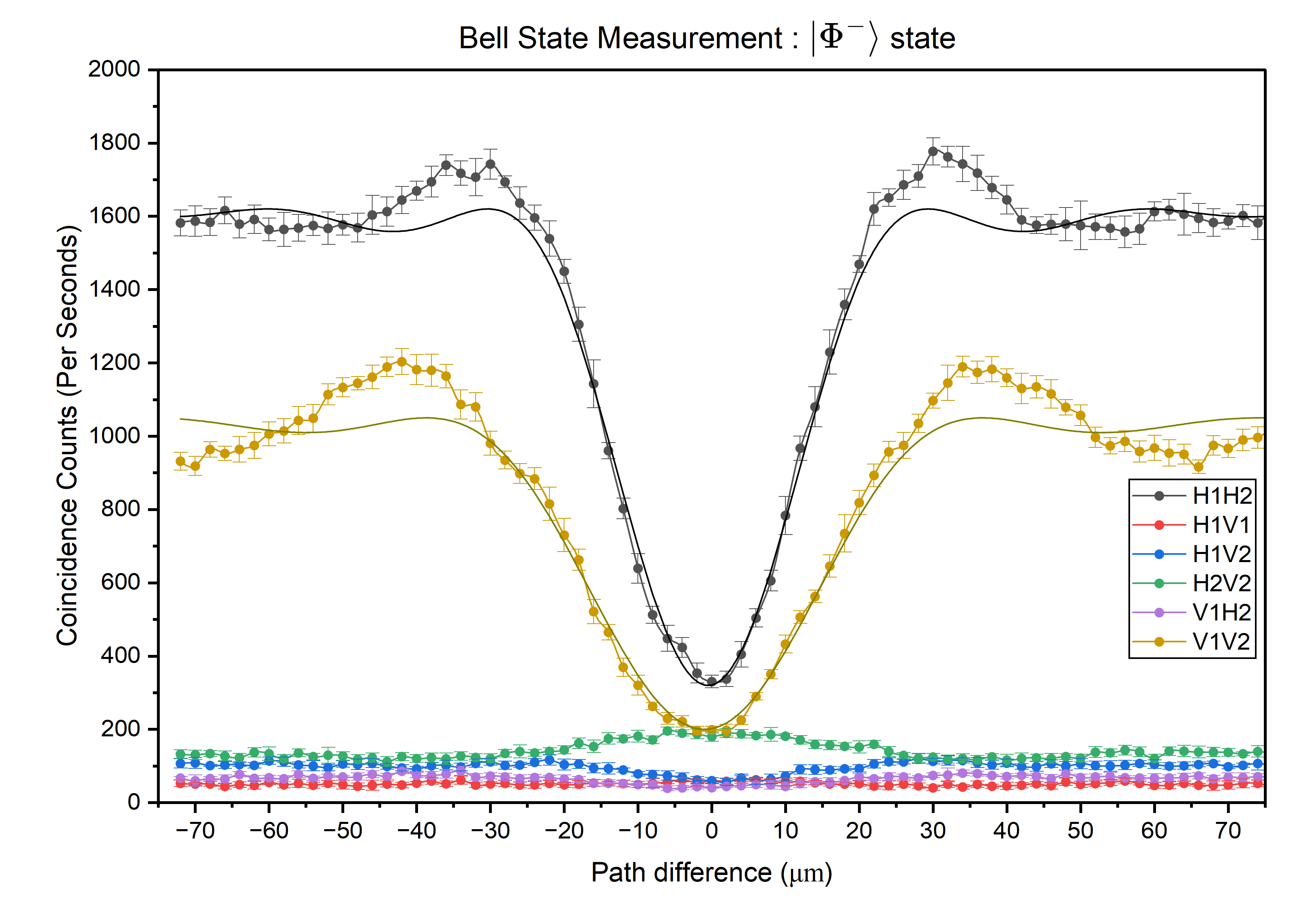} \\
(b)
\end{tabular}

\\[2mm]

\begin{tabular}{c}
\includegraphics[width=0.48\textwidth]{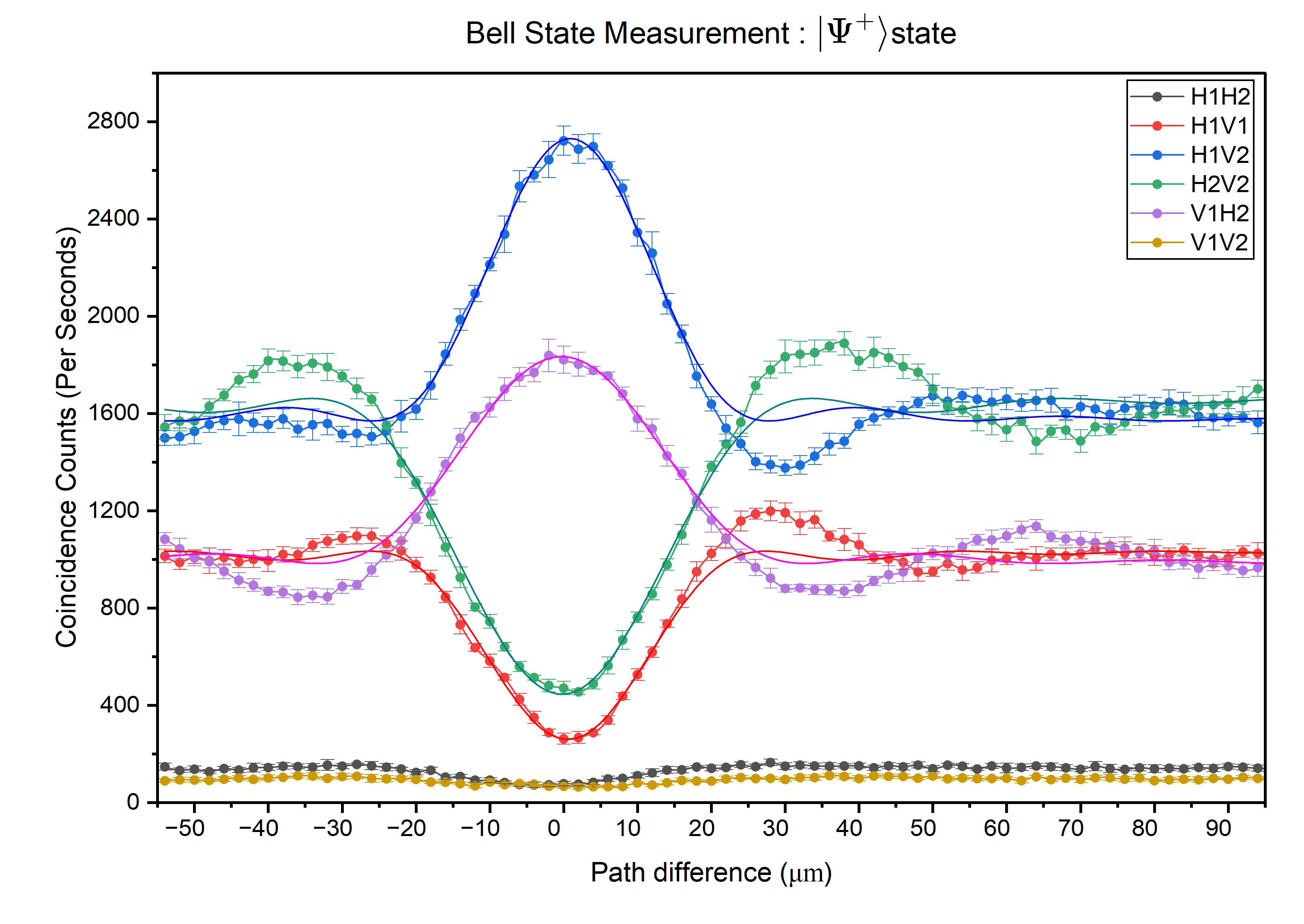} \\
(c)
\end{tabular}
&
\begin{tabular}{c}
\includegraphics[width=0.48\textwidth]{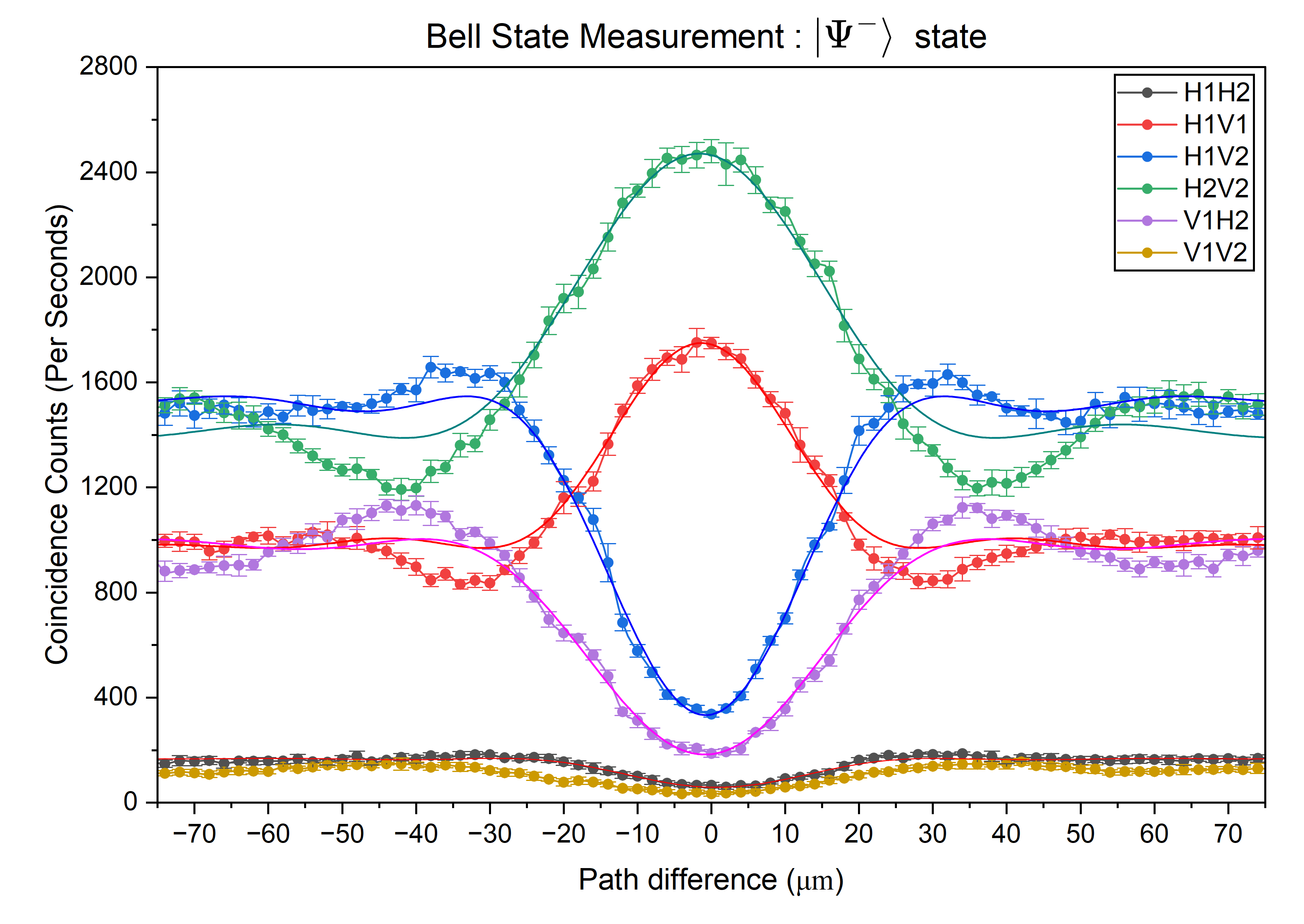} \\
(d)
\end{tabular}

\end{tabular}

\caption{
Bell-state measurement using a dielectric 50:50 beam splitter for different Bell states:
(a) $|\phi^+\rangle$ 
(b) $|\phi^-\rangle$ 
(c) $|\psi^+\rangle$ and
(d) $|\psi^-\rangle$.
}
\label{fig:BSM_images}
\end{figure*}

\begin{figure}[h]
    \centering
    \includegraphics[width=1\columnwidth]{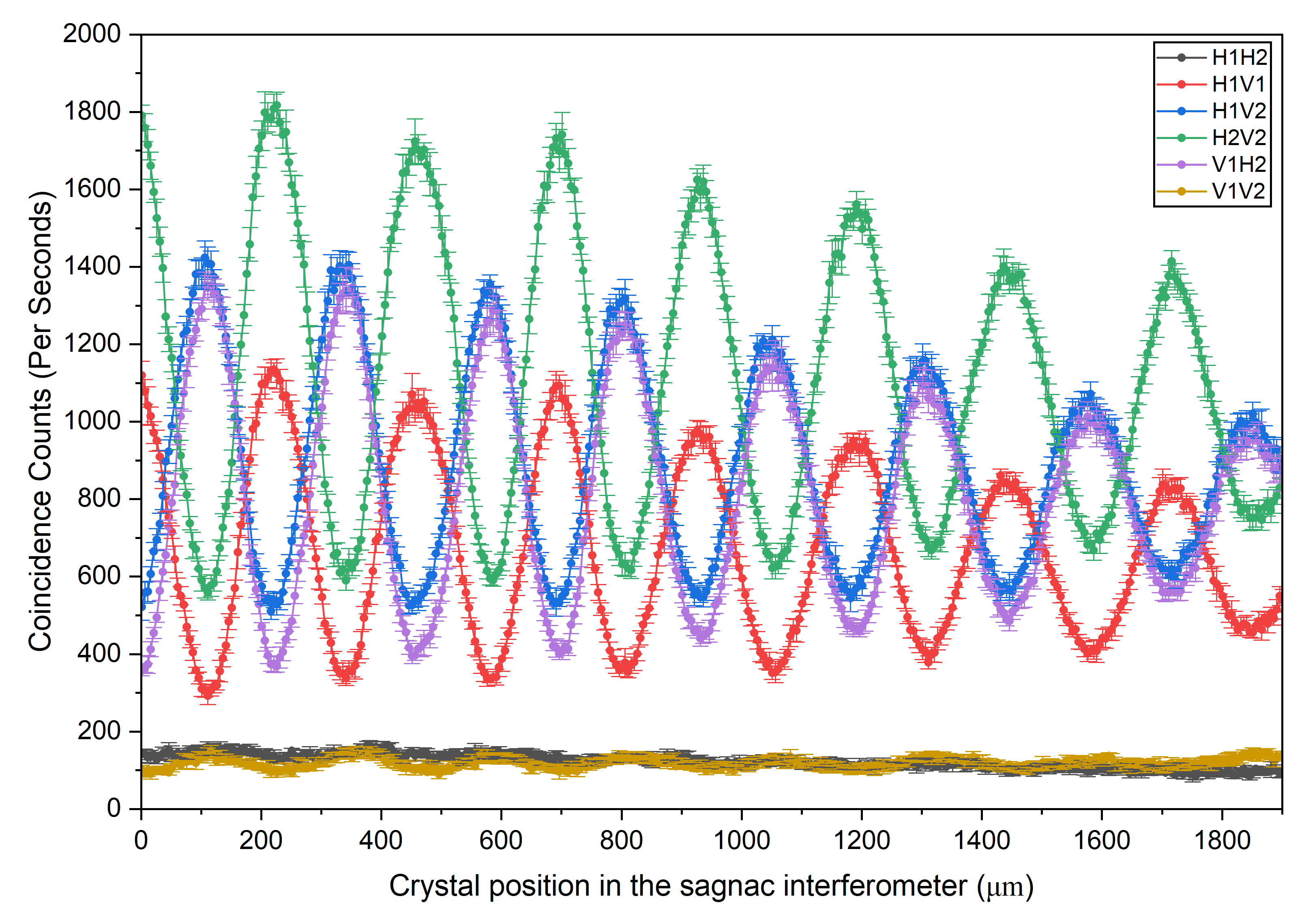}
    \caption{Effect of crystal translation on the Bell State Measurement ($|\psi^+\rangle$ to $|\psi^-\rangle$ state) .}\label{fig:crystal_translation_BSM}
\end{figure}

%%%%%%%%%%%%%%%%%%%%%%%%%%%%%%%

%In the previous section, the generated Bell states were verified with the complete quantum state tomography. 
The present section is dedicated to the discussion of the influence of the crystal translation on the Bell state measurement. The schematic diagram of the experimental setup is given in the earlier discussion. The half-wave plate in the idler arm is at $45^\circ$, now the generated quantum state of the signal-idler pair is given by, 
\begin{equation}
        |\psi\rangle = \frac{1}{\sqrt{2}} \left(|H_sV_i\rangle + e^{i \phi} |V_sH_i\rangle\right)
\end{equation}
As discussed earlier, the crystal translation from its balanced position leads to a continuous change in the relative phase $\phi$ and thus the generated quantum state. The crystal translation leads to the switch between the Bell states $|\psi^+\rangle$ and $|\psi^-\rangle$ in discrete intervals. The signal and idler photons are collected through single-mode fibers and polarization corrected before distributing it to a Bell state measurement setup containing a BS and PBS at each output port of BS.
\begin{figure}
\centering
\includegraphics[width=1\linewidth]{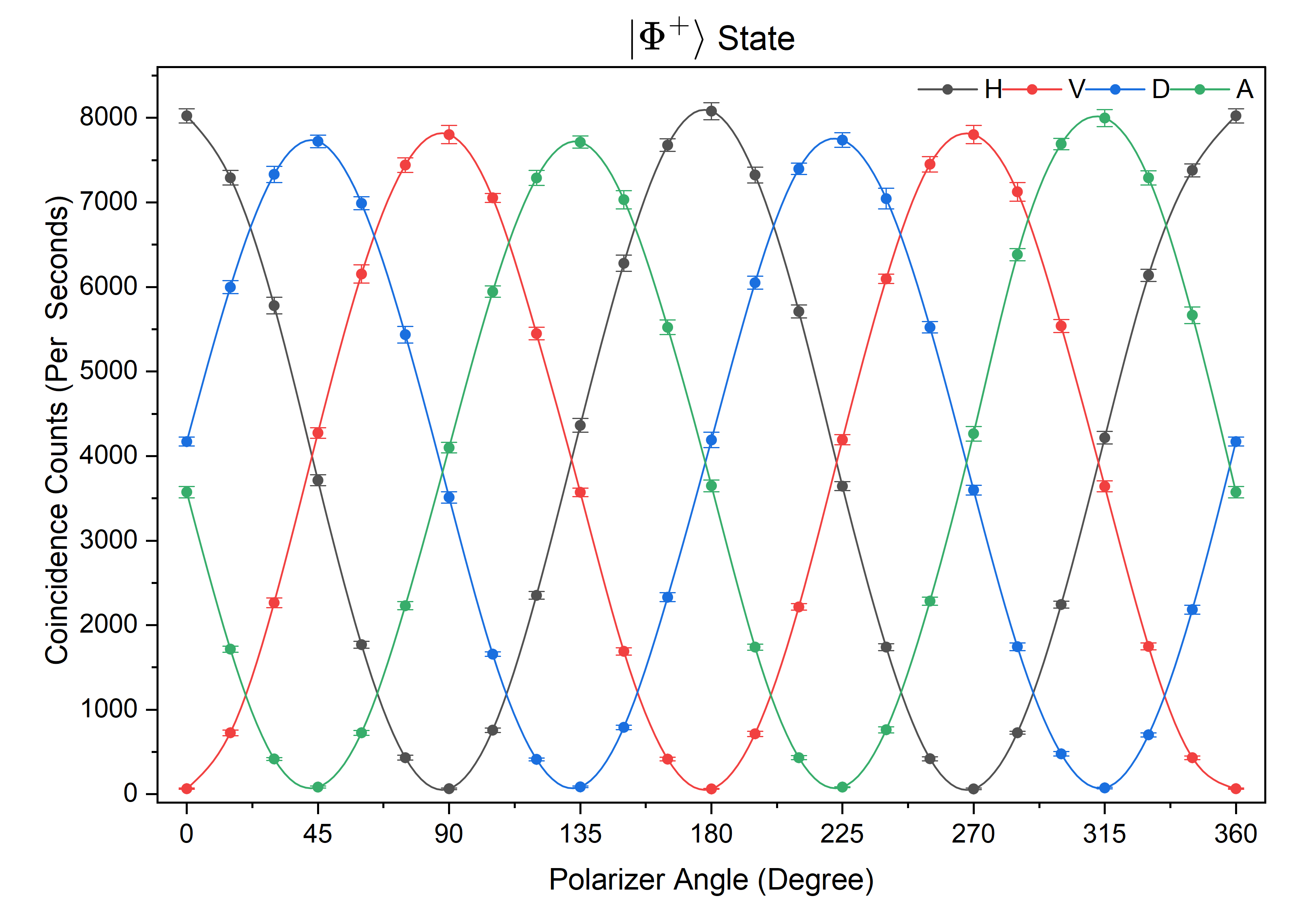}

\caption{
Correlation Curve for the $|\Phi^+\rangle$ state.
%(b) $|\Phi^-\rangle$ state
%(c) $|\Psi^+\rangle$ state
%(d) $|\Psi^-\rangle$ state
}
\label{fig:correlation curves}
\end{figure}

Before the Bell state measurement \cite{BSM_Weinfurter:1994xni}, we performed the Hong-Ou-Mandel interference experiment for all four Bell states. The theory suggests that the state $|\psi^-\rangle$ has the fermionic wavefunction and the remaining three Bell states have the Bosonic wavefunctions. Therefore, in the HOM interference experiment, the photon bunching is expected for the Bosonic states ($|\phi^\pm\rangle$ and $|\psi^+\rangle$ ) and the photon antibunching is expected for the Fermionic state  ($|\psi^-\rangle$) \cite{Hong1987,twophoton_interference_Pan2012}. The HOM interference is first measured with the dielectric beamsplitter. We have observed the HOM dip for the $|\phi^\pm\rangle$ states. But for the Bosonic state $|\psi^+\rangle$ the peak is observed instead of a dip. Similarly, for the Fermionic state $|\psi^-\rangle$, we observed the dip instead of the peak. It is due to the polarization dependence of the dielectric beamsplitter. These experimental results are in compliance with the results reported in the reference \cite{Ou2007}, as we have used dielectric beamsplitters for the Bell state measurement setup. The plot for the measured HOM interferences are given in Fig.~\ref{fig:HOMI_2x2_full}.  

We have repeated the HOM interference experiment with the polarization-maintaining fiber beamsplitter.  We have performed HOM interference for the Bell states $|\psi^+\rangle$ and $|\psi^-\rangle$.  These results comply with the theory, as we have observed a dip for the Bosonic state $|\psi^+\rangle$ and a peak for the Fermionic state $|\psi^-\rangle$. The plots for these experimental results are given in Fig.~\ref{fig:qutool's FBS HOM}.

We have performed the Bell-state measurement using a free-space dielectric BS by keeping PBSs at each output port of the BS. The experimental results of the BSM are given in Fig. \ref{fig:BSM_images}. The measured two-fold coincidence count rate between different output ports of the PBSs with respect to the crystal translation from its central position is given in the Fig.~\ref{fig:crystal_translation_BSM}. From the two-fold coincidence curves, it is clear that the generated quantum states are $|\psi^+\rangle$ and $|\psi^-\rangle$. We observed that, to generate the maximally entangled Bell state $|\psi^-\rangle$ from $|\psi^+\rangle$, the crystal needs to be translated by a distance $122\pm14~\mu m$. Therefore, every $122\pm14~\mu m$ distance, the generated states switches between $|\psi^+\rangle$ and $|\psi^-\rangle$. Since the influence of the crystal translation on the overlap between downconverted photon rings is minimal, the switching between $|\psi^+\rangle$ and $|\psi^-\rangle$ happens multiple times with minimal loss in the brightness of the entangled photon source. After translating the crystal over $1800~\mu m$ from the balanced position, the entanglement brightness drops by $21~\%$, which limits the translation distance on either side of the central position. This happens due to the noncollinear emission/geometry of the downconverted photons. Since the crystal translation leads to the misalignment for the downconversion angle $\theta_s=-\theta_i$.

\subsection{The complete quantum state tomography}

%%%%%%%%%%

%%%%%%%
\begin{figure}[t]
\centering

\setlength{\tabcolsep}{0pt}
\renewcommand{\arraystretch}{0.85}

\begin{tabular}{c@{}c@{}c}

% -------- Column headers --------
\textbf{State}& \textbf{Real $\bm{(\rho)}$} & \textbf{Imaginary $\bm{(\rho)}$} \\[0pt]

% -------- Row 1 --------
\raisebox{1.5cm}{\textbf{$\bm{|\Phi^{+}\rangle}$}} &
\includegraphics[width=0.44\columnwidth]{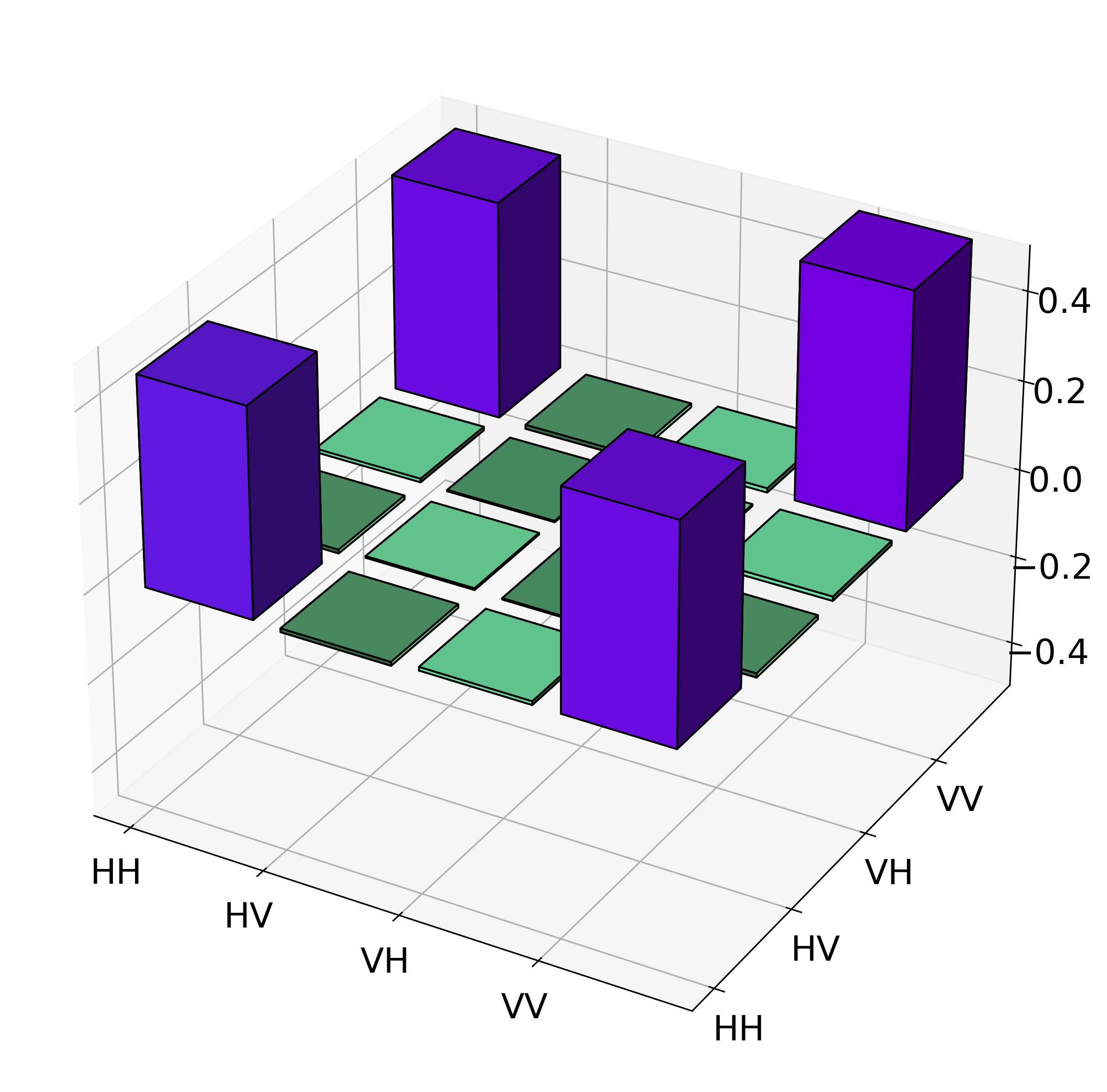} &
\includegraphics[width=0.44\columnwidth]{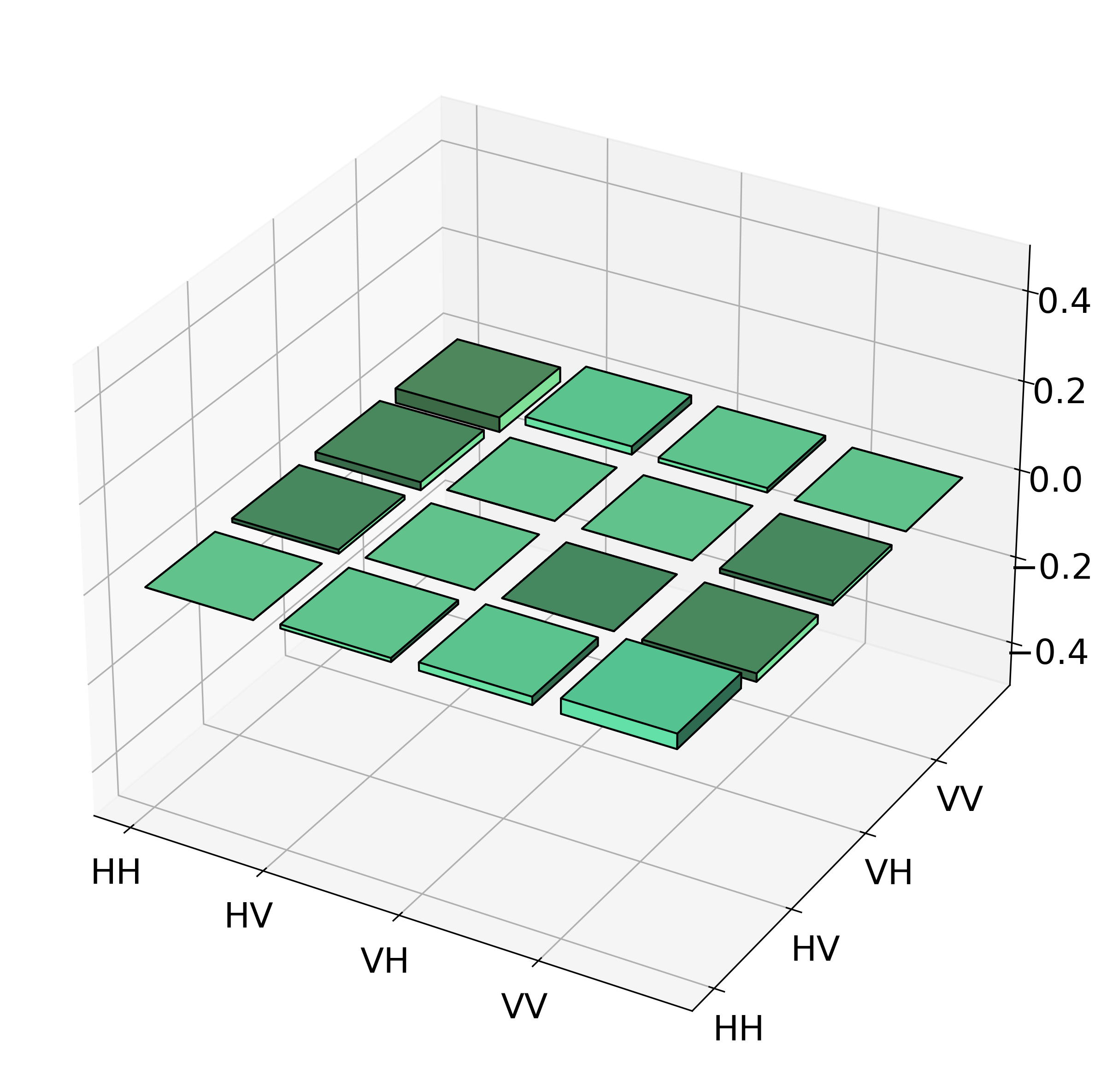} \\[-2pt]

% -------- Row 2 --------
\raisebox{1.5cm}{\textbf{$\bm{|\Phi^{-}\rangle}$}} &
\includegraphics[width=0.44\columnwidth]{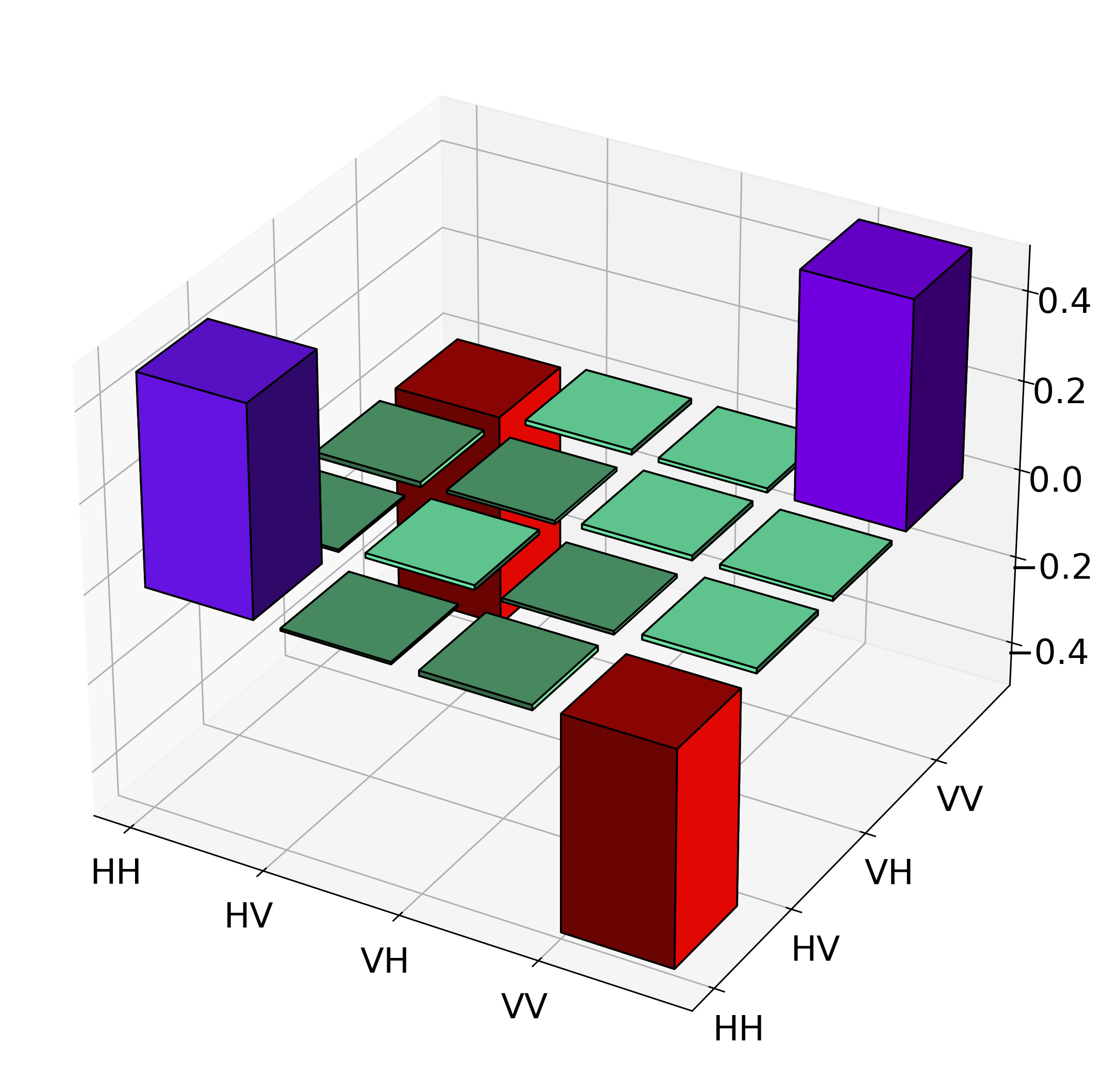} &
\includegraphics[width=0.44\columnwidth]{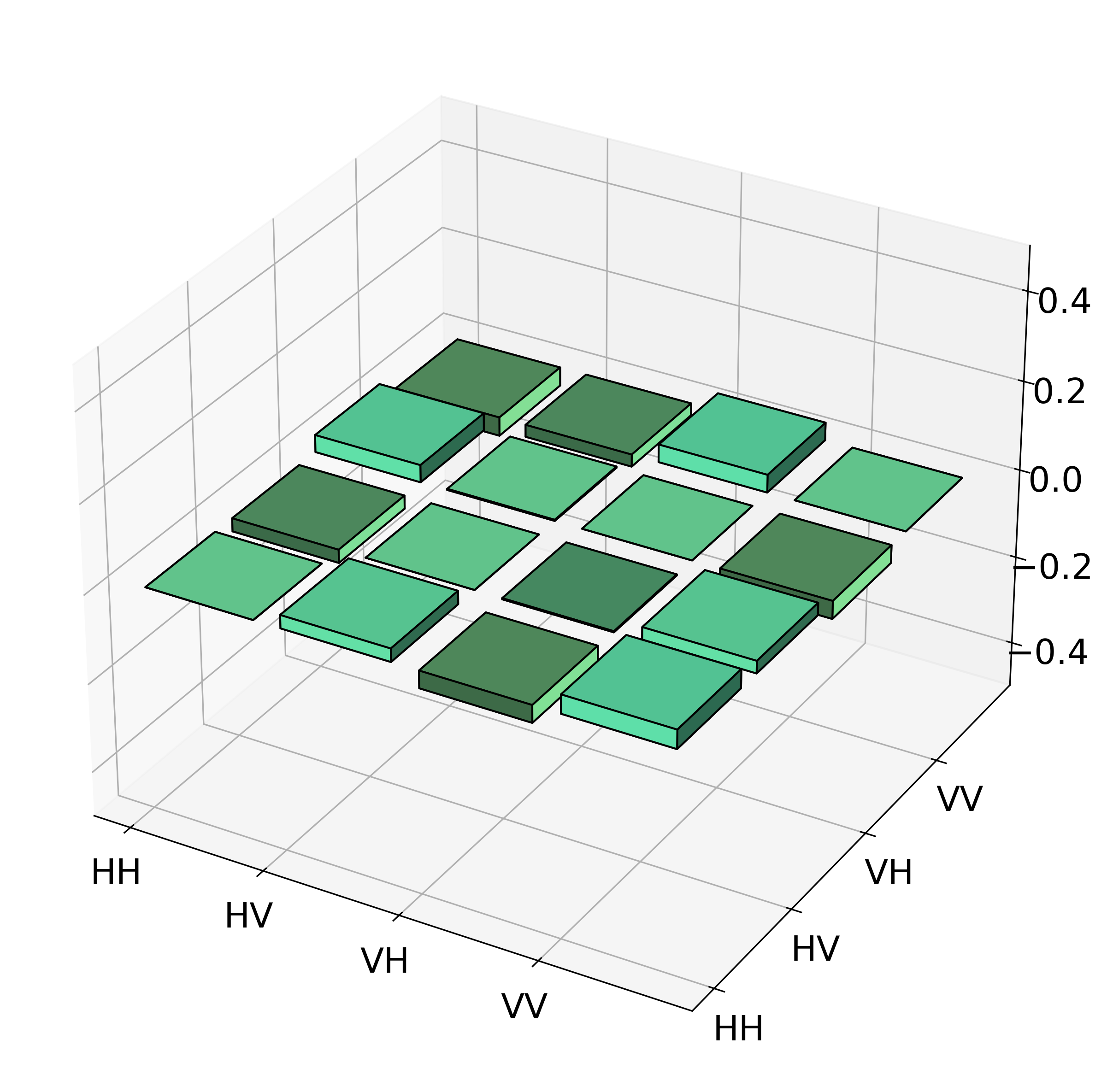} \\[-2pt]

% -------- Row 3 --------
\raisebox{1.5cm}{\textbf{$\bm{|\Psi^{+}\rangle}$}} &
\includegraphics[width=0.44\columnwidth]{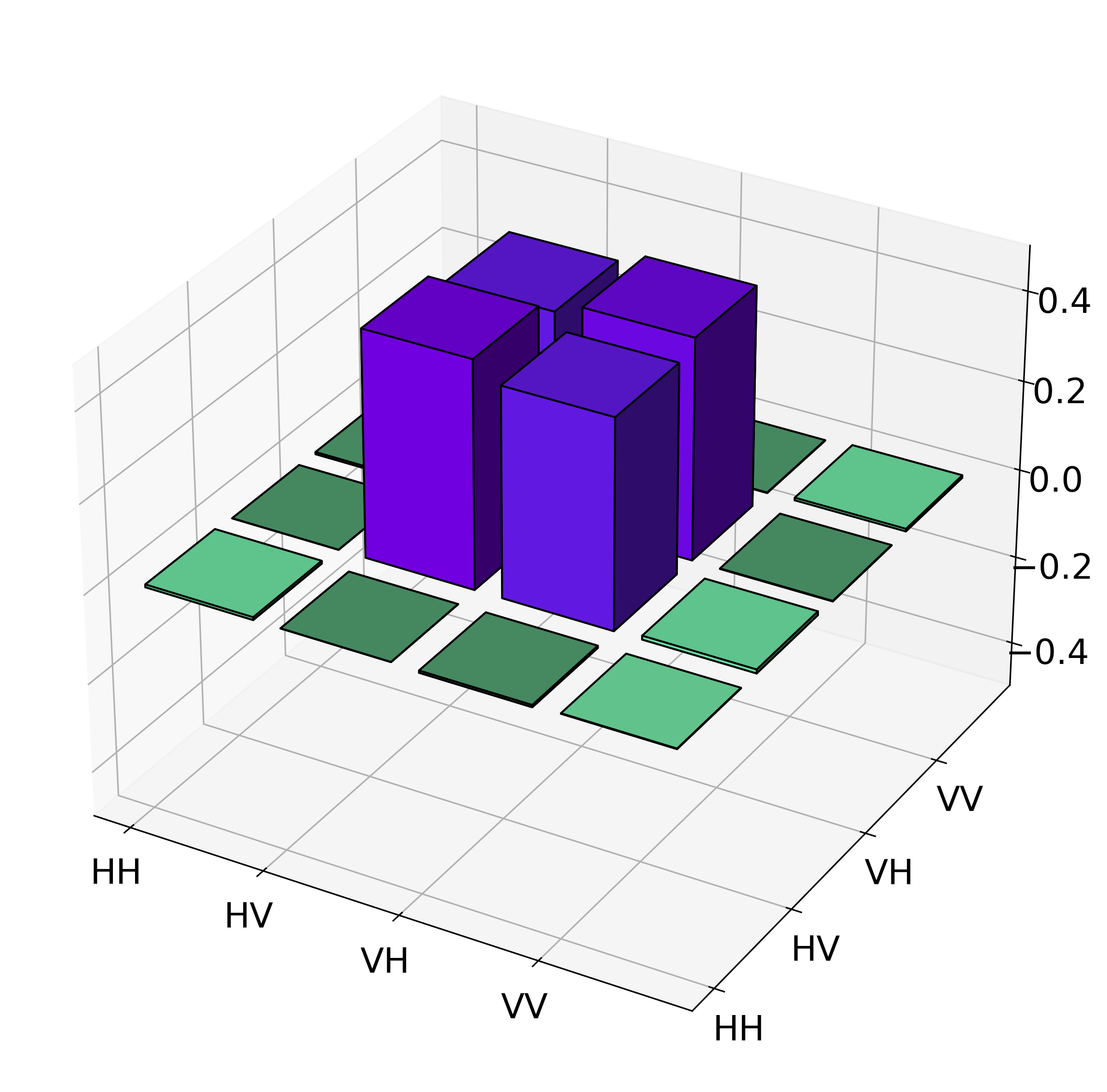} &
\includegraphics[width=0.44\columnwidth]{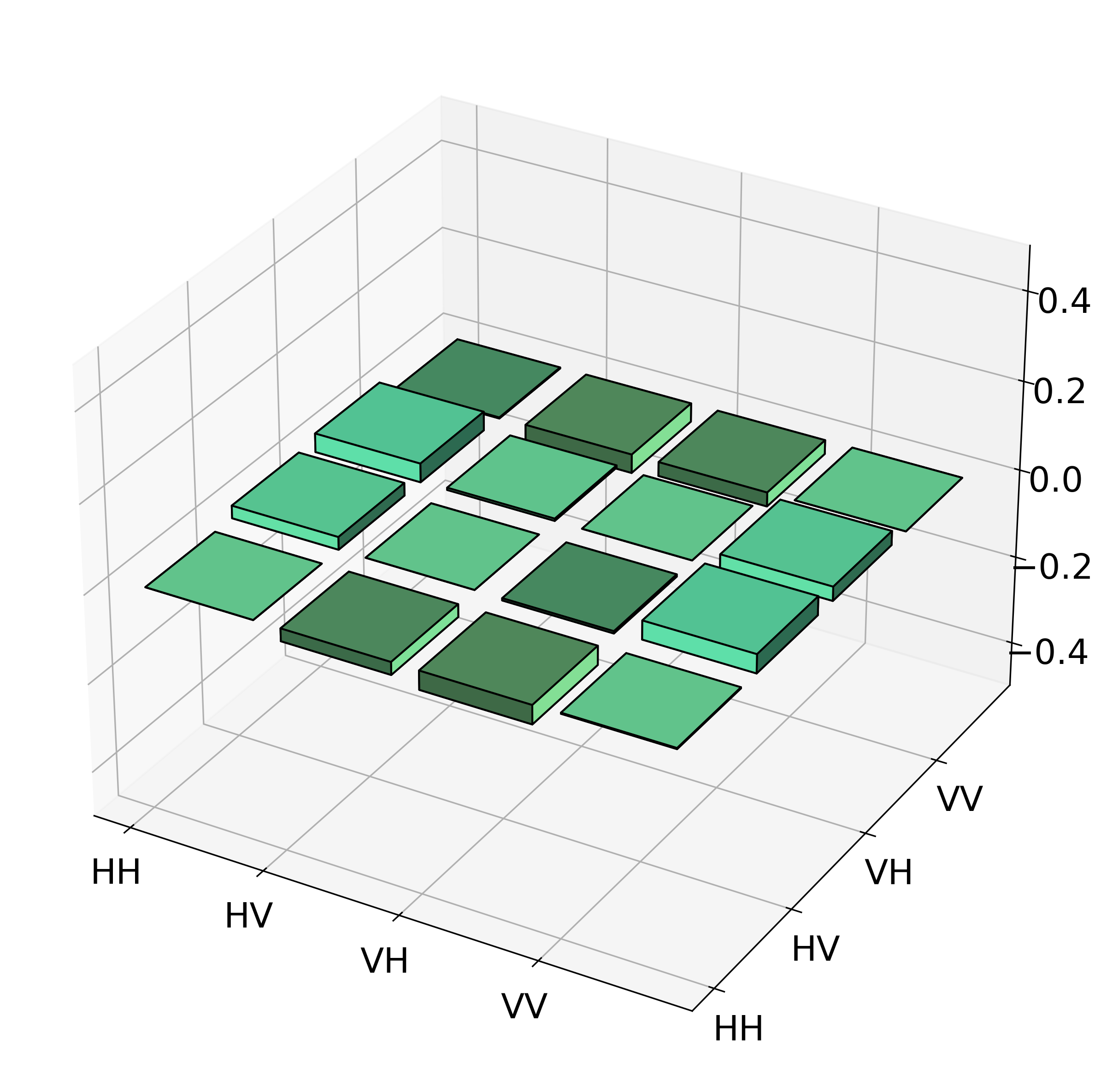} \\[-2pt]

% -------- Row 4 --------
\raisebox{1.5cm}{\textbf{$\bm{|\Psi^{-}\rangle}$}} &
\includegraphics[width=0.44\columnwidth]{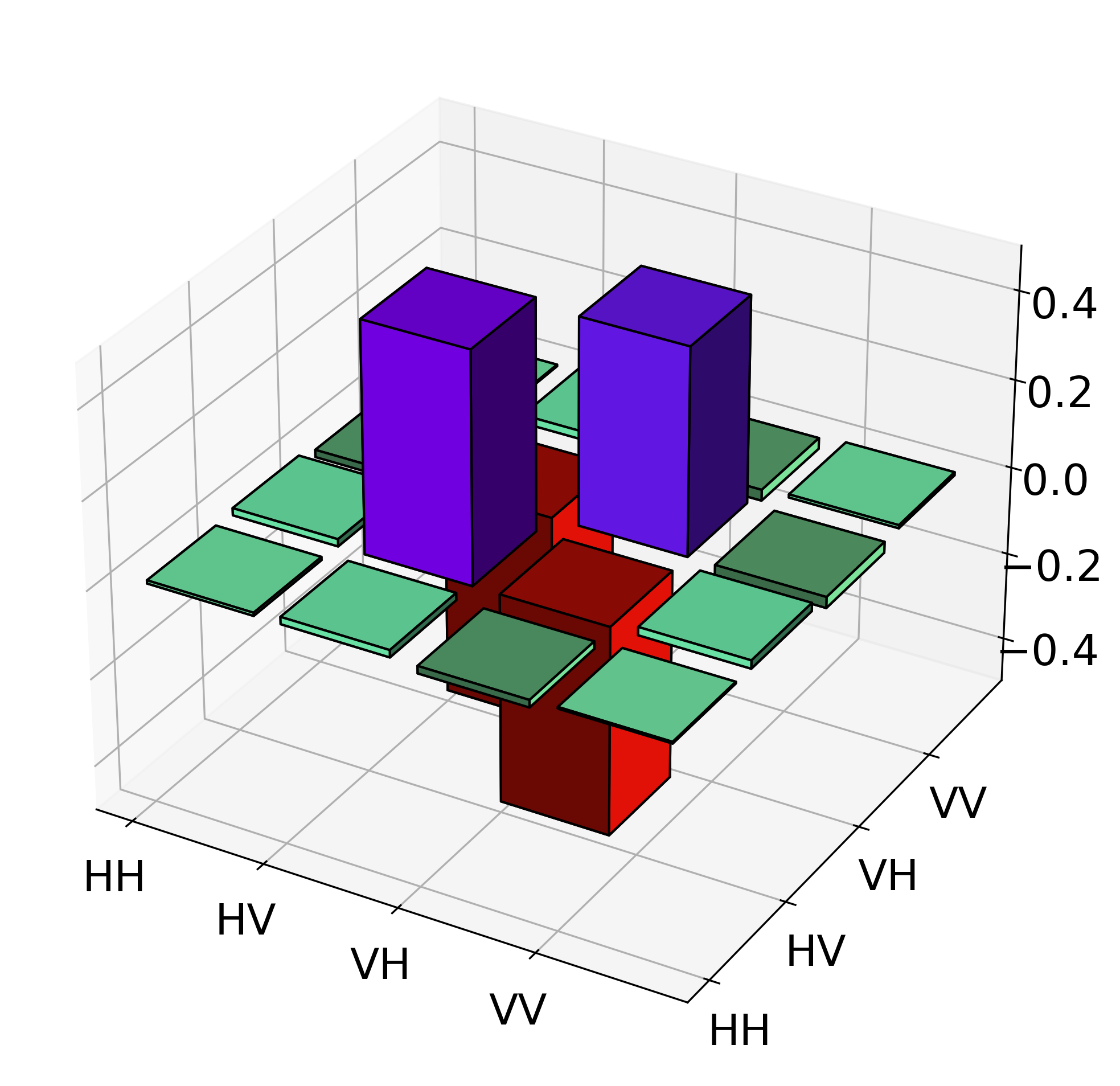} &
\includegraphics[width=0.44\columnwidth]{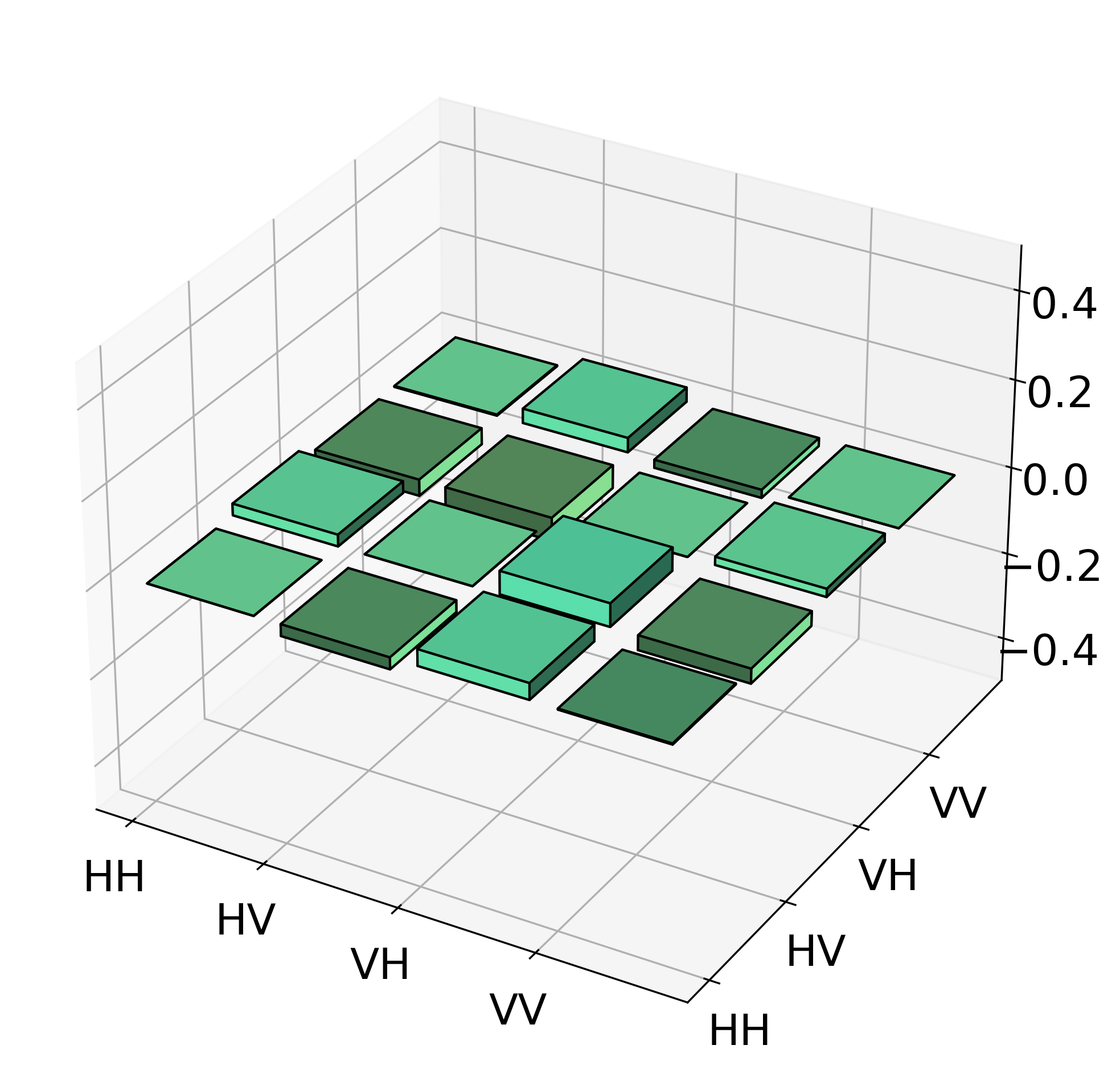} \\

% -------- Colorbar row --------
\multicolumn{3}{c}{
\includegraphics[width=0.8\columnwidth]{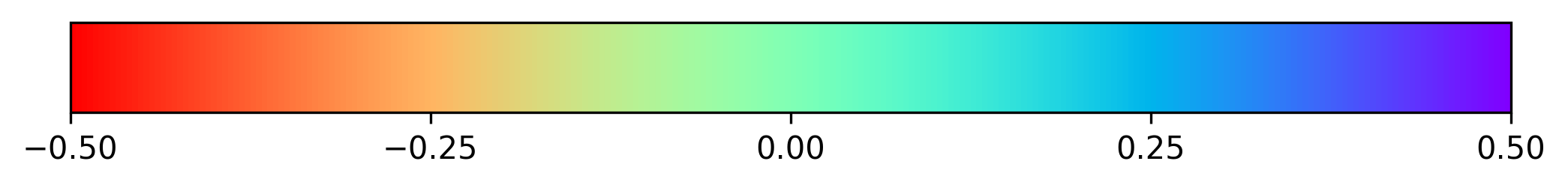}
}

\end{tabular}

\vspace{-8pt}

\caption{
Reconstructed density matrices for the four Bell states.
Left and right columns show the real and imaginary parts, respectively.
A single shared colorbar (bottom) ensures consistent scaling across all panels.
}

\label{fig:density_matrix}
\end{figure}

%%%%%%%

\begin{table*}[t]
\centering
\renewcommand{\arraystretch}{1.3}
\setlength{\tabcolsep}{14pt}
\begin{tabular}{c c c c c}

\hline\hline

 & $\bm{|\phi^+\rangle}$ & $\bm{|\phi^-\rangle}$ & $\bm{|\psi^+\rangle}$ & $\bm{|\psi^-\rangle}$ \\
\hline

\multicolumn{5}{c}{\textbf{Visibility in different basis (\%)}} \\
\hline

Computational basis (H/V) & 98.59 & 97.22 & 97.95 & 97.47 \\
Hadamard basis (+/-)     & 97.96 & 97.96 & 98.29 & 96.80 \\
Circular basis (R/L)     & 97.98 & 97.91 & 97.28 & 96.41 \\

\hline

\multicolumn{5}{c}{\textbf{State properties}} \\
\hline

CHSH  Inequality ($S$)         & 2.768 $\pm$ 0.011 & 2.749 $\pm$ 0.011 & 2.769 $\pm$ 0.011 & 2.748 $\pm$ 0.012 \\
Purity       & 0.9664 & 0.9651 & 0.9621 & 0.9536 \\
Fidelity     & 0.9796 & 0.9748 & 0.9743 & 0.9676 \\
Concurrence  & 0.9637 & 0.9654 & 0.9657 & 0.9528 \\

\hline\hline
\end{tabular}

\caption{Measured visibility in computational, hadamard, and circular basis, along with CHSH inequality ($S$-parameter), purity, fidelity, and concurrence for all four Bell states.}
\label{tab:EPS_parameters2}
\end{table*}

%%%%%%%%%%%%%

This section provides a detailed characterization of all four Bell states by performing complete quantum state tomography of the given state. The reconstructed density matrix offers the complete information about the generated quantum state, such as purity, fidelity, and concurrence. 

The quantum state tomography of the generated Bell states is performed by $16$ projective measurements for each state. The maximum likelihood estimation technique is used to find the closest possible density matrix to the reconstructed matrix \cite{MLE_James2001}. The density matrices of all four Bell states ($|\phi^\pm\rangle$, $|\psi^\pm\rangle$) are given in Fig. \ref{fig:density_matrix}. Further, the correlation curves for the $|\phi^+\rangle$ state are given in Fig.~\ref{fig:correlation curves}.  Along with the visibility in the different basis, the estimated purity, fidelity with respect to the actual quantum state and the concurrence of every Bell state are given in table \ref{tab:EPS_parameters2}. We have performed the Hanbury Brown and Twiss (HBT) experiment to measure the second order correlation function $(g^2(0))$. Experimentally measured $g^2(0)$ parameter is $0.034 \pm 0.010$. As given in table \ref{tab:EPS_parameters2}, the measured fidelity and purity are above $95\%$ for all four Bell states. The measured CHSH parameter is above $2.748\pm0.12$. 
%%%%%%%%%%%%%%%%%%%%%

\section{Conclusion}
We have thus realized an automated single monolithic downconverter that enables the generation of all four Bell states through crystal translation using a motorized stage and a rotation stage for the half-wave plate. We have also certified the quality of data through systematic measurements. Such a source is highly desirable for entanglement-based quantum communication, quantum networks, and photonic quantum computing. Our novel method demonstrates that the switching between different Bell states is possible without any additional optical components. This reduces the optical complexity of the source, which is not warranting additional optical components. Although there is a drop in the coincidence counts of $21 \%$  over a larger distance of $1800 ~ \mu m$, this effect is insignificant on the quality of the entanglement. Within the distance of $1800~\mu m$ from the balanced position, the flipping between the orthogonal Bell states is observed multiple times. Generating all four Bell states is crucial because they form a complete, orthonormal basis (the "Bell basis") for two-qubit systems, serving as the foundational building block for most quantum communication and computing protocols.
The controlled preparation and manipulation of the complete two-photon basis set $|\Phi^+\rangle$, $|\Phi^-\rangle$, $|\Psi^+\rangle$, and $|\Psi^-\rangle$ is important for multiple reasons. The realization of device-independent protocols relies on such a resource. For example, the ontological randomness of quantum random number generators is based on Bell state measurements. 

Furthermore, the security of entanglement-based protocols can be enhanced using these states. For example, in the Ekert protocol (E91), an additional layer of security can be provided by dynamically switching the source between these four Bell states. It is well known that the Bell states provide a complete basis for bi-partite states. This fact is exploited in protocols such as quantum teleportation and superdense coding. Entanglement Swapping is another protocol where our device would be useful. 

Additionally, testing and benchmarking of complex optical protocols can be facilitated by analyzing the transmission fidelity of each of the Bell states. The maximally entangled states help identify errors caused by noise or gate imperfections. Thus, a fully-automated device such as the one we have presented would be of substantial use for both basic studies and applications.\\

\begin{acknowledgments}
The authors acknowledge the support from DIAT(DU) under the grant-in-aid program. The authors acknowledge useful discussions with P. Kanaka Raju (School of Quantum Technology, DIAT, Pune, India). 
\end{acknowledgments}


%\nocite{*}

\bibliography{apssamp}% Produces the bibliography via BibTeX.

@PREAMBLE{
 "\providecommand{\noopsort}[1]{}" 
 # "\providecommand{\singleletter}[1]{#1}%" 
}

@article{Bennett_PhysRevLett.70.1895,
  title = {Teleporting an unknown quantum state via dual classical and Einstein-Podolsky-Rosen channels},
  author = {Bennett, Charles H. and Brassard, Gilles and Cr\'epeau, Claude and Jozsa, Richard and Peres, Asher and Wootters, William K.},
  journal = {Phys. Rev. Lett.},
  volume = {70},
  issue = {13},
  pages = {1895--1899},
  numpages = {0},
  year = {1993},
  month = {Mar},
  publisher = {American Physical Society},
  doi = {10.1103/PhysRevLett.70.1895},
  url = {https://link.aps.org/doi/10.1103/PhysRevLett.70.1895}
}

@book{Nielsen_Chuang_2010, 
place={Cambridge}, 
title={Quantum Computation and Quantum Information: 10th Anniversary Edition},
publisher={Cambridge University Press}, 
author={Nielsen, Michael A. and Chuang, Isaac L.}, 
year={2010}}

@article{HOM_PhysRevLett.59.2044,
  title = {Measurement of subpicosecond time intervals between two photons by interference},
  author = {Hong, C. K. and Ou, Z. Y. and Mandel, L.},
  journal = {Phys. Rev. Lett.},
  volume = {59},
  issue = {18},
  pages = {2044--2046},
  numpages = {0},
  year = {1987},
  month = {Nov},
  publisher = {American Physical Society},
  doi = {10.1103/PhysRevLett.59.2044},
  url = {https://link.aps.org/doi/10.1103/PhysRevLett.59.2044}
}

@article{Pan_Entang_swap_PhysRevLett.80.3891,
  title = {Experimental Entanglement Swapping: Entangling Photons That Never Interacted},
  author = {Pan, Jian-Wei and Bouwmeester, Dik and Weinfurter, Harald and Zeilinger, Anton},
  journal = {Phys. Rev. Lett.},
  volume = {80},
  issue = {18},
  pages = {3891--3894},
  numpages = {0},
  year = {1998},
  month = {May},
  publisher = {American Physical Society},
  doi = {10.1103/PhysRevLett.80.3891},
  url = {https://link.aps.org/doi/10.1103/PhysRevLett.80.3891}
}

@article{MDI-QKD_PhysRevLett.108.130503,
  title = {Measurement-Device-Independent Quantum Key Distribution},
  author = {Lo, Hoi-Kwong and Curty, Marcos and Qi, Bing},
  journal = {Phys. Rev. Lett.},
  volume = {108},
  issue = {13},
  pages = {130503},
  numpages = {5},
  year = {2012},
  month = {Mar},
  publisher = {American Physical Society},
  doi = {10.1103/PhysRevLett.108.130503},
  url = {https://link.aps.org/doi/10.1103/PhysRevLett.108.130503}
}

@article{SPDC_PhysRevLett.75.4337,
  title = {New High-Intensity Source of Polarization-Entangled Photon Pairs},
  author = {Kwiat, Paul G. and Mattle, Klaus and Weinfurter, Harald and Zeilinger, Anton and Sergienko, Alexander V. and Shih, Yanhua},
  journal = {Phys. Rev. Lett.},
  volume = {75},
  issue = {24},
  pages = {4337--4341},
  numpages = {0},
  year = {1995},
  month = {Dec},
  publisher = {American Physical Society},
  doi = {10.1103/PhysRevLett.75.4337},
  url = {https://link.aps.org/doi/10.1103/PhysRevLett.75.4337}
}

@article{anwarAli_10.1063/5.0023103,
    author = {Anwar, Ali and Perumangatt, Chithrabhanu and Steinlechner, Fabian and Jennewein, Thomas and Ling, Alexander},
    title = {Entangled photon-pair sources based on three-wave mixing in bulk crystals},
    journal = {Review of Scientific Instruments},
    volume = {92},
    number = {4},
    pages = {041101},
    year = {2021},
    month = {04},
    issn = {0034-6748},
    doi = {10.1063/5.0023103},
    url = {https://doi.org/10.1063/5.0023103},
   }

@article{Guo_2017,
doi = {10.7567/APEX.10.062801},
url = {https://doi.org/10.7567/APEX.10.062801},
year = {2017},
month = {may},
publisher = {The Japan Society of Applied Physics},
volume = {10},
number = {6},
pages = {062801},
author = {Guo, Kai and Christensen, Erik N. and Christensen, Jesper B. and Koefoed, Jacob G. and Bacco, Davide and Ding, Yunhong and Ou, Haiyan and Rottwitt, Karsten},
title = {High coincidence-to-accidental ratio continuous-wave photon-pair generation in a grating-coupled silicon strip waveguide},
journal = {Applied Physics Express},

}

@article{Kwait_99_PhysRevA.60.R773,
  title = {Ultrabright source of polarization-entangled photons},
  author = {Kwiat, Paul G. and Waks, Edo and White, Andrew G. and Appelbaum, Ian and Eberhard, Philippe H.},
  journal = {Phys. Rev. A},
  volume = {60},
  issue = {2},
  pages = {R773--R776},
  numpages = {0},
  year = {1999},
  month = {Aug},
  publisher = {American Physical Society},
  doi = {10.1103/PhysRevA.60.R773},
  url = {https://link.aps.org/doi/10.1103/PhysRevA.60.R773}
}

@article{Marco_F_Fiorentino:07,
author = {Marco Fiorentino and Sean M. Spillane and Raymond G. Beausoleil and Tony D. Roberts and Philip Battle and Mark W. Munro},
journal = {Opt. Express},
number = {12},
pages = {7479--7488},
publisher = {Optica Publishing Group},
title = {Spontaneous parametric down-conversion in periodically poled KTP waveguides and bulk crystals},
volume = {15},
month = {Jun},
year = {2007},
url = {https://opg.optica.org/oe/abstract.cfm?URI=oe-15-12-7479},
doi = {10.1364/OE.15.007479},
}

@article{Konig_10.1063/1.1668320,
    author = {König, Friedrich and Wong, Franco N. C.},
    title = {Extended phase matching of second-harmonic generation in periodically poled KTiOPO4 with zero group-velocity mismatch},
    journal = {Applied Physics Letters},
    volume = {84},
    number = {10},
    pages = {1644-1646},
    year = {2004},
    month = {03},
    issn = {0003-6951},
    doi = {10.1063/1.1668320},
    
}

@article{Tanzilli,
    author = {S. Tanzilli and  W. Tittel and H. De Riedmatten and H. Zbinden and P. Baldi,  M. DeMicheli and D.B. Ostrowsky and N. Gisin},
    title = {PPLN waveguide for quantum communication},
    journal = {The European Physical Journal D - Atomic, Molecular, Optical and Plasma Physics},
    volume = {18},
    number = {10},
    pages = {155-160},
    year = {2002},
    month = {02},
    issn = {1434-6079},
    doi = {10.1140/epjd/e20020019},
    
}

@article{Kim_PhysRevA.73.012316,
  title = {Phase-stable source of polarization-entangled photons using a polarization Sagnac interferometer},
  author = {Kim, Taehyun and Fiorentino, Marco and Wong, Franco N. C.},
  journal = {Phys. Rev. A},
  volume = {73},
  issue = {1},
  pages = {012316},
  numpages = {5},
  year = {2006},
  month = {Jan},
  publisher = {American Physical Society},
  doi = {10.1103/PhysRevA.73.012316},
  url = {https://link.aps.org/doi/10.1103/PhysRevA.73.012316}
}

@article{Sundar:25_VariableBellState,
author = {S. Sundar and M. V. Jabir and D. Razansky},
journal = {J. Opt. Soc. Am. B},
number = {8},
pages = {1764--1770},
publisher = {Optica Publishing Group},
title = {Variable Bell states in a polarization-entangled photon source},
volume = {42},
month = {Aug},
year = {2025},
doi = {10.1364/JOSAB.560031},

}

@article{teleportation_DAurelio,
author = {D’Aurelio and E. Simone and Bayerbach and J. Matthias and Barz and Stefanie},
journal = {npj Quantum Information},
number = {8},
pages = {37},
title = {Boosted quantum teleportation},
volume = {11},
month = {March},
year = {2025},
doi = {10.1038/s41534-025-00992-4},

}

@article{bellStateMeasurement,
  author  = {Bianchi, Luca and Marconi, Carlo and Bacco, Davide},
  title   = {Bell state measurements in quantum optics: a review of recent progress and open challenges},
  journal = {arXiv preprint arXiv:2509.18756},
  year    = {2025},
  doi     = {10.48550/arXiv.2509.18756}
}

@article{Kim_PhysRevLett.86.1370,
  title = {Quantum Teleportation of a Polarization State with a Complete Bell State Measurement},
  author = {Kim, Yoon-Ho and Kulik, Sergei P. and Shih, Yanhua},
  journal = {Phys. Rev. Lett.},
  volume = {86},
  issue = {7},
  pages = {1370--1373},
  numpages = {0},
  year = {2001},
  month = {Feb},
  publisher = {American Physical Society},
  doi = {10.1103/PhysRevLett.86.1370},
  url = {https://link.aps.org/doi/10.1103/PhysRevLett.86.1370}
}

@book{Ou2007,
  author    = {Zhe-Yu Jeff Ou},
  title     = {Multi-Photon Quantum Interference},
  publisher = {Springer},
  address   = {New York},
  year      = {2007},
  chapter      = {4, Sec. 4.1}
}

@article{Jabir2017,
  author  = {M. V. Jabir and G. K. Samanta},
  title   = {Robust, high brightness, degenerate entangled photon source at room temperature},
  journal = {Sci. Rep.},
  volume  = {7},
  number  = {1},
  pages   = {12613},
  year    = {2017},
  doi     = {10.1038/s41598-017-12709-5}
}

@article{PPLN_Szlachetka2023,
  author  = {J. Szlachetka and K. Joarder and P. Kolenderski},
  title   = {Ultrabright source of non-degenerate polarization-entangled photon pairs based on off-the-shelf polarization optics},
  journal = {Appl. Phys. Lett.},
  volume  = {123},
  number  = {14},
  pages   = {144001},
  year    = {2023},
  doi     = {10.1063/5.0159000}
}

@article{BBO_SansaPerna2022,
  author  = {A. Sansa Perna and E. Ortega and M. Gr{\"a}fe and F. Steinlechner},
  title   = {Visible-wavelength polarization-entangled photon source for quantum communication and imaging},
  journal = {Appl. Phys. Lett.},
  volume  = {120},
  number  = {7},
  pages   = {074001},
  year    = {2022},
  doi     = {10.1063/5.0069992}
}

@article{BiBO_Rangarajan2009,
  author  = {R. Rangarajan and M. Goggin and P. G. Kwiat},
  title   = {Optimizing type-I polarization-entangled photons},
  journal = {Opt. Express},
  volume  = {17},
  number  = {21},
  pages   = {18920--18933},
  year    = {2009},
  doi     = {10.1364/OE.17.018920}
}

@article{PPKTP_linear_Park2025,
  author  = {K. Park and J. Lee and D.-G. Im and D. Kim and Y. S. Ihn},
  title   = {Ultrabright fiber-coupled polarization-entangled photon source with spectral brightness surpassing 2.0\,MHz\,mW$^{-1}$\,nm$^{-1}$},
  journal = {Adv. Photonics Res.},
  volume  = {6},
  number  = {10},
  pages   = {2500024},
  year    = {2025},
  doi     = {10.1002/adpr.202500024}
}

@article{PPKTP_linear_Guo2023,
  author  = {S. Guo and K. Shang},
  title   = {High-flux, high-visibility entangled photon source obtained with a non-collinear type-II {PPKTP} crystal pumped by a broadband continuous-wave diode laser},
  journal = {Opt. Commun.},
  volume  = {545},
  pages   = {129586},
  year    = {2023},
  doi     = {10.1016/j.optcom.2023.129586}
}

@article{sagnac_Cai2022,
  author  = {N. Cai and W.-H. Cai and S. Wang and F. Li and R. Shimizu and R.-B. Jin},
  title   = {Broadband-laser-diode pumped periodically poled potassium titanyl phosphate-Sagnac polarization-entangled photon source},
  journal = {J. Opt. Soc. Am. B},
  volume  = {39},
  number  = {1},
  pages   = {77--82},
  year    = {2022},
  doi     = {10.1364/JOSAB.437808}
}

@article{BSM_Weinfurter:1994xni,
    author = "Weinfurter, H.",
    title = "{Experimental Bell-State Analysis}",
    doi = "10.1209/0295-5075/25/8/001",
    journal = "EPL",
    volume = "25",
    number = "8",
    pages = "559",
    year = "1994"
}

@article{Hong1987,
  author  = {C. K. Hong and Z. Y. Ou and L. Mandel},
  title   = {Measurement of subpicosecond time intervals between two photons by interference},
  journal = {Phys. Rev. Lett.},
  volume  = {59},
  number  = {18},
  pages   = {2044--2046},
  year    = {1987},
  doi     = {10.1103/PhysRevLett.59.2044}
}

@article{twophoton_interference_Pan2012,
  author  = {J.-W. Pan and Z.-B. Chen and C.-Y. Lu and H. Weinfurter and A. Zeilinger and M. {\.Z}ukowski},
  title   = {Multiphoton entanglement and interferometry},
  journal = {Rev. Mod. Phys.},
  volume  = {84},
  number  = {2},
  pages   = {777--838},
  year    = {2012},
  doi     = {10.1103/RevModPhys.84.777}
}

@article{MLE_James2001,
  author  = {D. F. V. James and P. G. Kwiat and W. J. Munro and A. G. White},
  title   = {Measurement of qubits},
  journal = {Phys. Rev. A},
  volume  = {64},
  number  = {5},
  pages   = {052312},
  year    = {2001},
  doi     = {10.1103/PhysRevA.64.052312}
}

@article{Jin2014,
  author  = {R.-B. Jin and R. Shimizu and K. Wakui and H. Benichi and M. Sasaki},
  title   = {Widely tunable high-quality photon-pair source at telecom wavelengths using a Sagnac interferometer},
  journal = {Opt. Express},
  volume  = {22},
  pages   = {11498},
  year    = {2014}
}

@article{Mishra2024,
  author  = {Sarika Mishra and R. P. Singh},
  title   = {Transformation of Bell states using linear optics},
  journal = {Phys. Open},
  volume  = {18},
  pages   = {100199},
  year    = {2024},
  doi     = {10.1016/j.physo.2023.100199}
}

@article{Kim2019,
  author  = {Heonoh Kim and Osung Kwon and Han Seb Moon},
  title   = {Pulsed Sagnac source of polarization-entangled photon pairs in telecommunication band},
  journal = {Sci. Rep.},
  volume  = {9},
  pages   = {5031},
  year    = {2019},
  doi     = {10.1038/s41598-019-41633-z}
}

@article{Gisin1998,
  author  = {N. Gisin and H. Bechmann-Pasquinucci},
  title   = {Bell inequality, Bell states and maximally entangled states for n qubits},
  journal = {Phys. Lett. A},
  volume  = {246},
  pages   = {1--6},
  year    = {1998},
  doi     = {10.1016/S0375-9601(98)00516-7}
}

@article{Shih1988,
  author  = {Y. H. Shih and C. O. Alley},
  title   = {New Type of Einstein-Podolsky-Rosen-Bohm Experiment Using Pairs of Light Quanta Produced by Optical Parametric Down Conversion},
  journal = {Phys. Rev. Lett.},
  volume  = {61},
  pages   = {2921--2924},
  year    = {1988},
  doi     = {10.1103/PhysRevLett.61.2921}
}

@article{Ekert1991,
  author  = {A. K. Ekert},
  title   = {Quantum cryptography based on Bell's theorem},
  journal = {Phys. Rev. Lett.},
  volume  = {67},
  pages   = {661--663},
  year    = {1991},
  doi     = {10.1103/PhysRevLett.67.661}
}
%\includegraphics{fig_1.eps}

%%%%%%%%%%%%%%%%%%%%%%%%%%%%%%%%%%%%%%%%%%%
%include EMCCD images (Done)
% include temp vs wavelength
%include Hong-ou dip for each bell state, CHSS (Done)
%power vs coincidence, accidental coincidence
% include tomography results, what are visibility %values for each bell state (Done)

\end{document}